%% file: onepoint.tex
\documentclass[12pt, a4paper, fleqn]{article}

\usepackage[nottoc]{tocbibind} 

\RequirePackage[l2tabu, orthodox]{nag}
\usepackage{silence}
\ActivateWarningFilters[pdftoc]

\usepackage[T1]{fontenc}
\usepackage[utf8]{inputenc}

\usepackage[top=20mm, bottom=20mm, left=25mm, right=25mm]{geometry}

\usepackage{array,longtable}
\usepackage{multirow}

\usepackage[table]{xcolor}

\usepackage{amsmath, amssymb, amsfonts, amsthm}
\usepackage{centernot}
\usepackage{bm}  
\usepackage{mathtools} 
\usepackage{graphicx}

\usepackage[colorlinks=true, linktoc=all, linkcolor=black, citecolor=red, urlcolor=blue, backref=page]{hyperref}

\usepackage{etoolbox}
\makeatletter
\patchcmd{\BR@backref}{\newblock}{\newblock(}{}{}
\patchcmd{\BR@backref}{\par}{)\par}{}{}
\makeatother

\numberwithin{equation}{section}
\usepackage{tikz}
\usetikzlibrary{calc,decorations.pathmorphing}
\usetikzlibrary{decorations.pathreplacing}
\usetikzlibrary{decorations.markings, arrows.meta}
\usetikzlibrary{shapes}
\usepackage{pgf}
\usetikzlibrary{calc}

\newcommand{\tore}{
\coordinate (base) at (0, -.2);
\draw[dashed] (-1, -1) rectangle (1, 1);
\node[fill, circle, minimum size = 2mm, inner sep = 0] at (0, 0) {};
}

\usetikzlibrary{shapes.misc}
\tikzset{cross/.style={cross out, draw=black, thick, fill=none, minimum size=2*(#1-\pgflinewidth), inner sep=0pt, outer sep=0pt}, cross/.default={3pt}}

\newcommand{\vertices}{
  \coordinate (base) at (0, 1);
  \draw (0, 0) node[fill, circle, minimum size = 2mm, inner sep = 0]{};
  \draw (0, 3) node[cross]{};
  \draw (3, 0) node[cross]{};
  \draw (3, 3) node[cross]{};
}

\tikzset{
  loops/.style={line width = 2pt, blue, rounded corners=6pt, opacity=0.55}
}

\allowdisplaybreaks

\begin{document}

\section*{}
\begin{flushleft}
  {\bfseries\sffamily\Large 
    Torus one-point functions in critical loop models
    \vspace{1.5cm}
    \\
    \hrule height .6mm
  }
  \vspace{1.5cm}

  {\bfseries\sffamily 
    Paul Roux, Sylvain Ribault, Jesper Lykke Jacobsen
  }
  \vspace{4mm}

 {\textit{\!\!
(PR, SR, JJ) Institut de physique théorique, CEA, CNRS,
Université Paris-Saclay \vspace{2mm}
\\
(PR, JJ) Laboratoire de Physique de l'École Normale Supérieure, ENS, Université PSL, \\ \hspace{1.65cm}  CNRS, Sorbonne Université, Université Paris Cité, Paris, France
}}

  \vspace{4mm}

  {\textit{E-mail:} \texttt{
      paul.roux@phys.ens.fr,
      sylvain.ribault@ipht.fr,
      jesper.jacobsen@phys.ens.fr
    }}
\end{flushleft}
\vspace{7mm}

{\noindent\textsc{Abstract:}
  \vspace{2mm}

  \noindent
We show that in critical loop models, torus 1-point functions can be expressed in terms of sphere 4-point functions at a different central charge. Unlike in the Moore--Seiberg formalism, crossing symmetry on the sphere therefore implies modular covariance on the torus. 

We systematically compute torus 1-point functions in critical loop models, using a numerical
bootstrap approach. We focus on the 1-point functions of the 6 simplest primary fields, which
give rise to 10 solutions of modular covariance equations. Such 1-point functions are infinite
linear combinations of conformal blocks. The coefficients are products of double Gamma
functions, times polynomial functions of loop weights. For each solution, we determine the first
6 to 12 polynomials.  }

\clearpage

\hrule 
\tableofcontents
\vspace{5mm}
\hrule
\vspace{5mm}
\hypersetup{linkcolor=blue!70!black}

\section{Introduction}

\subsubsection{The challenge of solving critical loop models}

Critical loop models are a class of 2d conformal field theories, which describe the critical limits of statistical models of non-intersecting loops on 2d lattices, such as the $O(n)$ and $Q$-state Potts models (see \cite{rib24} for a review). Critical loop models were claimed to be exactly solvable, based on the determination of 235 examples of structure constants for 4-point functions on the sphere \cite{nrj23}. To solve these models, we would need to determine all structure constants for all correlation functions. However, unlike in simpler theories such as minimal models or Liouville theory, 4-point structure constants do not factorize into 3-point structure constants. So our full (but conjectural) knowledge of 3-point structure contants \cite{jnrr25} is not enough for solving the models.

In order to understand structure constants, it is helpful to have systematic results. However,
since sphere 4-point functions depend on 4 fields, studying them systematically involves large
numbers of cases. In this article, we focus on torus 1-point functions, which depend on only 1
field. Loop models contain infinitely many primary fields $V_{(r,s)}$, where the first Kac index $r\in \frac12\mathbb{N}$ is half the number of legs: we have studied all
independent non-vanishing torus 1-point functions $\left<V_{(r_1,s_1)}\right>$ with $r_1\leq 3$. On
the sphere, known results do not even cover all 4-point functions of fields with $r\leq
1$. (Compare the table of cases \eqref{tab} with the corresponding table of 4-point functions
\cite[Section 1.3]{nrj23}.)

In critical loop models, a correlation function is not fully determined by a Riemann surface and
$N$ fields: we also need to specify the connectivity pattern between the fields. This pattern is encoded in a combinatorial map \cite{gjnrs23}, and in Section \ref{subsec: maptor} we will study the combinatorial maps that are relevant to torus 1-point functions. Then correlation functions are computed by solving
conformal bootstrap equations: crossing symmetry on the sphere, modular covariance on a
torus. Which solution corresponds to which combinatorial map is known in some but not all
cases. The favourable cases are when the combinatorial map is characterized by its intersections
with topologically nontrivial cycles. On the torus, all cycles are combinatorially equivalent,
allowing us to characterize very few combinatorial maps. To circumvent this difficulty, we will
use the sphere-torus relation.

\subsubsection{Help from the sphere-torus relation}

It is well-known that 1-point Virasoro blocks on the torus can be expressed in terms of 4-point Virasoro blocks on the sphere \cite{flno09}. But we need more: a relation between torus 1-point functions and sphere 4-point functions. We will find such a relation, based on relations for the technical ingredients of correlation functions:
\begin{itemize}
 \item combinatorial maps \eqref{phi},
 \item full (non-chiral) conformal blocks, including logarithmic blocks \eqref{gag},
 \item structure constants \eqref{dpd}.
\end{itemize}
In particular, the relation for combinatorial maps will solve our problem of characterizing
1-point maps on the torus. This is because on a sphere with 4 punctures, we have 3
combinatorially different cycles, corresponding to the $s$, $t$ and $u$ channels. This allows us
to associate 3 intersection numbers to any given combinatorial map. These numbers are not enough
to characterize arbitrary 4-point maps on the sphere, but they do characterize the maps that come from the torus, modulo parity.

The sphere-torus relation works for correlation functions not only in critical loop models, but also in other solvable CFTs such as Liouville theory and minimal models. This seems to contradict well-known statements that modular covariance on the torus does not follow from crossing symmetry on the sphere \cite{ms89b}. And there are known examples of CFTs that are consistent on the sphere, but not modular covariant. In Section \ref{mccs} we will discuss under which assumptions modular covariance does follow from crossing symmetry.

\subsubsection{Primary fields on the torus and on the sphere}

Let us introduce the primary fields that appear in critical loop models, and whose correlation functions we want to compute. We use the following notations for the central charge $c$, conformal dimension $\Delta$, momentum $P$, and Kac indices $(r,s)$:
\begin{align}
 c= 13-6\beta^2-6\beta^{-2} \quad ,\quad \Delta = \frac{c-1}{24}+P^2 \quad ,\quad P_{(r,s)} = \frac12\left(\beta r-\beta^{-1}s\right)\ .
 \label{cdp}
\end{align}
We define degenerate, diagonal and non-diagonal fields by their left and right momentums $(P,\bar P)$, plus the condition that degenerate fields have vanishing singular vectors:
\begin{align}
\renewcommand{\arraystretch}{1.5}
 \begin{array}{|r|l|l|l|l|}
  \hline
  \text{Name} & \text{Notation} & \text{Conditions} & (P,\bar P) & \text{Spin}
  \\
  \hline \hline
  \text{Degenerate} &  V^d_{\langle 1,s\rangle} &  s\in\mathbb{N}^* & \left(P_{(1,s)},P_{(1,s)}\right) & 0
  \\
  \hline
  \text{Diagonal} & V_P & P\in\mathbb{C} & \left(P, P\right) & 0
  \\
  \hline
  \text{Non-diagonal} & V_{(r,s)} &
   r\in\frac12\mathbb{N}^*, \ s\in\frac{1}{r}\mathbb{Z} & \left(P_{(r,s)},P_{(-r,s)}\right) & rs
  \\
  \hline
 \end{array}
 \label{fields}
\end{align}
We define a spectrum as a set of primary fields that appear in a given channel for a given correlation function. For example, the $s$-channel spectrum of a 4-point function. When decomposing torus 1-point functions into conformal blocks, the relevant spectrums are subsets of
\begin{align}
  \mathcal{S}=
 \left\{V_{(r,s)}\right\}_{\substack{r\in \frac12\mathbb{N}\\ s\in\frac{1}{r}\mathbb{Z}}} \ ,
 \label{spec}
\end{align}
where by convention
\begin{align}
 \left. \frac{1}{r}\mathbb{Z} \right|_{r=0} = s_0+2\mathbb{Z}\qquad , \qquad V_{(0,s)} = V_{P_{(0,s)}}\ ,
 \label{conv}
\end{align}
where $s_0 \in \mathbb{C}$ is arbitrary.
The contractible loop weight $n$ of the loop model is related to
the central charge of the CFT by
\begin{align}
 n = -2\cos(\pi\beta^2) \ .
 \label{nb}
\end{align}
A non-diagonal primary field $V_{(r,s)}$ corresponds to a puncture with $2r$ legs.
A diagonal primary field $V_P$ corresponds to a puncture that changes the weight of loops to
\begin{align}
 w(P) = 2\cos(2\pi \beta P) \ .
 \label{wP}
\end{align}
In particular, the primary field $V_0\equiv V_{(0,\frac12)}$ corresponds to a puncture that gives
weight $w=0$ to loops around it. This primary field plays an important role in the sphere-torus
relation, which relates a torus 1-point function $\left<V_1\right>$ to a sphere 4-point function
of the type $\left<V'_0V'_1V'_0V'_0\right>^\text{sphere}$, where the primes indicate that the variables should be transformed according to the sphere-torus relation which we are about to display, for example $\beta' = \frac{\beta}{\sqrt{2}}$.
(We put the field $V_1'$ in second position for later convenience.)
A torus 1-point conformal block or a sphere 4-point conformal block also depend
on a channel field $V$, and the sphere-torus relation is not the same for channel fields as for
the external field $V_1$:
\begin{align}
\label{rel}
\renewcommand{\arraystretch}{1.3}
  \begin{array}{| r | r | c  c |}
    \hline
    \multicolumn{2}{|c|}{} & \text{Torus} & \text{Sphere}
    \\ \hline \multicolumn{2}{|c|}{\text{Correlation function}} & \left<V_1\right> & \left<V'_0V_1'V'_0V'_0\right>
     \\ \hline \multicolumn{2}{|c|}{\text{Conformal block}} &
     \begin{tikzpicture}[baseline = (base), scale = .3]
   \coordinate (base) at (0, 0);
   \node[below] at (-2.3, 0) {$V_1$};
   \node[right] at (3, 0) {$V$};
   \draw[ultra thick] (-2.5, 0) to (0, 0);
  \draw[ultra thick] plot [smooth, tension = .8] coordinates {(0, 0) (2, 1) (3, 0) (2, -1) (0, 0)};
 \end{tikzpicture}
 & \begin{tikzpicture}[baseline = (base), scale = .25]
   \coordinate (base) at (0, 0);
  \draw[ultra thick] (-1 ,2) -- (0, 0) -- (4, 0) -- (5, 2);
  \draw[ultra thick] (-1, -2) -- (0, 0);
  \draw[ultra thick] (5, -2) -- (4, 0);
  \node[left] at (-1, -2) {$V'_0$};
  \node[right] at (5,2) {$V'_0$};
  \node[right] at (5,-2) {$V'_0$};
  \node[left] at (-.8,2) {$V_1'$};
  \node[below] at (2, 0) {$V'$};
 \end{tikzpicture}
 \\ \hline \multicolumn{2}{|c|}{\text{Modulus}} & \tau & \tau
     \\  \multicolumn{2}{|c|}{\text{Nome}} & q & \sqrt{q}
    \\ \hline \multicolumn{2}{|c|}{\text{Central charge parameter}} & \beta & \frac{\beta}{\sqrt2}
     \\  \multicolumn{2}{|c|}{\text{Loop weight}} & n & -\sqrt{2-n}
    \\ \hline
    \multirow{4}{*}{\text{External field}} & \text{Momentum} & P_1 & \frac{P_1}{\sqrt2} \\
    & \text{Kac indices}& (r_1, s_1) & (r_1, \frac{s_1}{2})
    \\  & \text{Weight} & w_1 & \sqrt{w_1+2}
    \\  & \text{Spin} & S_1 & \frac{S_1}{2}
    \\ \hline
    \multirow{4}{*}{\text{Channel field}} & \text{Momentum} & P & \sqrt2 P \\
    & \text{Kac indices} & (r, s) & (2r, s)
     \\  & \text{Weight} & w & w
    \\  & \text{Spin} & S & 2S
    \\ \hline
  \end{array}
\end{align}
For example,
\begin{subequations}
\begin{align}
 P'_{(2r,s)} = P_{(2r,s)}(\beta') = \sqrt{2} P_{(r,s)}(\beta) \quad , \quad P'_{(r_1,\frac{s_1}{2})} = P_{(r_1,\frac{s_1}{2})}(\beta') = \tfrac{1}{\sqrt{2}} P_{(r_1,s_1)}(\beta)\ ,
\end{align}
\begin{align}
 \text{External field: } w'_1 = w(P'_1,\beta')=w\left(\tfrac{P_1}{\sqrt{2}},\tfrac{\beta}{\sqrt{2}}\right)=\sqrt{w_1+2}\ ,
\end{align}
\begin{align}
 \text{Channel field: } w' = w(P',\beta')=w\left(\sqrt{2}P,\tfrac{\beta}{\sqrt{2}}\right)=w\ .
\end{align}
\end{subequations}

\subsubsection{Torus 1-point structure constants}

Our main goal is to determine the structure constants that appear in the decomposition of a torus 1-point function into conformal blocks:
\begin{align}
 \left<V_{(r_1,s_1)}\right> = \sum_{(r,s)\in \mathcal{S}} D_{(r,s)}\mathcal{G}_{(r,s)} \ .
 \label{vdg}
\end{align}
Here the spectrum $\mathcal{S}$ is given in Eq. \eqref{spec}. The conformal blocks
$\mathcal{G}_{(r,s)}$ \eqref{grs} are universal, known functions that depend on the
modulus $\tau$ of the torus, whereas the structure constants $D_{(r,s)}$ do not depend on
$\tau$. Inspired by similar results for sphere 4-point functions \cite[(1.11)]{nrj23}, we
conjecture that the structure constants are of the form
\begin{align}
\label{eq:15}
  D_{(r, s)} = \frac{\hat D_{(r, s)}}{\kappa_{(r, s)}} \left( \frac{d_{(r, s)}}{r} +\delta_{r\in\mathbb{N}^*}\delta_{s\in\mathbb{Z}} \frac{ \Theta_1^{r,s}}{w - w_{(r, s)}} \right)\ .
\end{align}
In this expression, the reference structure constant $\hat D_{(r, s)}$ is deduced from the 3-point structure constant \cite[(1.7)]{jnrr25} in Eq. \eqref{hd}, and computed explicitly as a combination of double Gamma functions in Eq. \eqref{hdrs}. The quantities $\kappa_{(r, s)}$ \eqref{kappa} and $\Theta_1^{r,s}$ \eqref{dors} are known combinations of trigonometric functions, with $\kappa_{(r,s)}$ already playing the same role for 4-point functions on the sphere.
It remains to determine
$d_{(r, s)}$, which depends on the choice of a combinatorial map, parametrized by 3 integers $(m,n,p)$ with $r_1=m+n+p$.
We conjecture that $d_{(r,s)}$ is polynomial in
\begin{itemize}
 \item the contractible loop weight $n$,
 \item the channel loop weight $w$ if the channel spectrum includes diagonal fields, i.e.\ if the combinatorial map is $(r_1,0,0)$,
 \item the external loop weight $w_1$ if the external field $V_1$ is diagonal, i.e.\ if the combinatorial map is $(0,0,0)$.
\end{itemize}
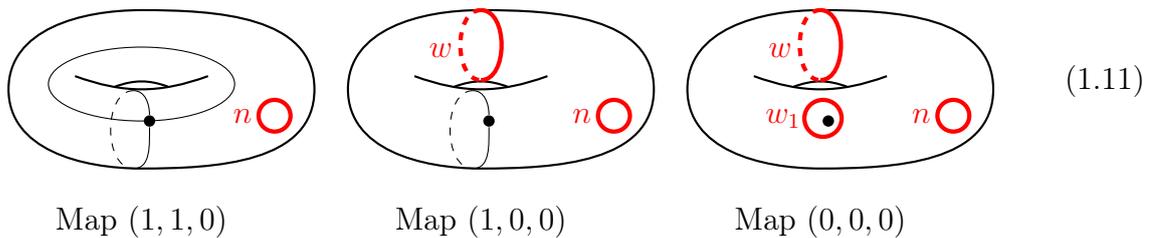
\begin{align}
 \begin{tikzpicture}[baseline=(base), scale = .35]
 \coordinate (base) at (0, 0);
 \draw[thick] (-2.5, .5) to [out = -20, in = -160] (2.5, .5);
  \draw[thick] (-1, .1) to [out = 20, in = 160] (1, .1);
  \draw[thick] (-5, 0) to [out = 90, in = 180] (0, 3) to [out = 0, in = 90] (6.5, 0) to [out = -90, in = 0] (0, -3) to [out = 180, in = -90] (-5, 0);
  \node[fill, circle, minimum size = 1.5mm, inner sep = 0] at (0.3, -1.2) (A1) {};
  \draw[out = 90, in = 0] (A1) to (-.3, -.05);
  \draw[out = -90, in =0] (A1) to (-.3, -3);
  \draw[dashed, out = 180, in = 180] (-.3, -.05) to (-.3, -3);
  \draw (0, .2) ellipse (3.5cm and 1.4cm);
  \draw[ultra thick, red] (5, -1) circle (.6);
  \node[red] at (3.8, -1) {$n$};
  \node at (0, -5){Map $(1,1,0)$};
\end{tikzpicture}
\quad
\begin{tikzpicture}[baseline=(base), scale = .35]
 \coordinate (base) at (0, 0);
 \draw[thick] (-2.5, .5) to [out = -20, in = -160] (2.5, .5);
  \draw[thick] (-1, .1) to [out = 20, in = 160] (1, .1);
  \draw[thick] (-5, 0) to [out = 90, in = 180] (0, 3) to [out = 0, in = 90] (6.5, 0) to [out = -90, in = 0] (0, -3) to [out = 180, in = -90] (-5, 0);
  \node[fill, circle, minimum size = 1.5mm, inner sep = 0] at (0.3, -1.2) (A1) {};
  \draw[out = 90, in = 0] (A1) to (-.3, -.05);
  \draw[out = -90, in =0] (A1) to (-.3, -3);
  \draw[dashed, out = 180, in = 180] (-.3, -.05) to (-.3, -3);
  \draw[ultra thick, red] (0, 3) to [out =0, in = 0] (0, .35);
  \draw[ultra thick, dashed, red]  (0, 3) to [out =180, in = 180] (0, .35);
  \draw[ultra thick, red] (5, -1) circle (.6);
  \node[red] at (3.8, -1) {$n$};
  \node[red] at (-1.5, 1.5) {$w$};
  \node at (0, -5){Map $(1, 0,0)$};
\end{tikzpicture}
\quad
\begin{tikzpicture}[baseline=(base), scale = .35]
 \coordinate (base) at (0, 0);
 \draw[thick] (-2.5, .5) to [out = -20, in = -160] (2.5, .5);
  \draw[thick] (-1, .1) to [out = 20, in = 160] (1, .1);
  \draw[thick] (-5, 0) to [out = 90, in = 180] (0, 3) to [out = 0, in = 90] (6.5, 0) to [out = -90, in = 0] (0, -3) to [out = 180, in = -90] (-5, 0);
  \node[fill, circle, minimum size = 1.5mm, inner sep = 0] at (0.3, -1.2) (A1) {};
 \draw[ultra thick, red] (0, 3) to [out =0, in = 0] (0, .35);
  \draw[ultra thick, dashed, red]  (0, 3) to [out =180, in = 180] (0, .35);
  \draw[ultra thick, red] (5, -1) circle (.6);
  \draw[ultra thick, red] (.1, -1.1) circle (.7);
  \node[red] at (3.8, -1) {$n$};
  \node[red] at (-1.4, -1.1){$w_1$};
  \node[red] at (-1.5, 1.5) {$w$};
  \node at (0, -5){Map $(0,0,0)$};
\end{tikzpicture}
\end{align}
Moreover, our numerical results suggest that for
$r\geq \frac12$ the degree of $d_{(r,s)}$ is
\begin{subequations}
  \begin{align}
    \label{eq:18}
    \deg_{w_1} d_{(r, s)} &= \left\lfloor r - \tfrac12 \right\rfloor\ , & & \text{ for } \left\langle V_{P_1} \right\rangle\ , \\
    \deg_w d_{(r, s)} &= 2r-r_1\ ,  & & \text{ for the map }(r_1,0,0)\ ,  \\
    \deg_n d_{(r, s)} &= \left\lfloor \left(r-\tfrac12\right)^2\right\rfloor \ ,  & & =\deg_n \kappa_{(r, s)}\ , \text{ see Eq. }   \eqref{degk}\ .
  \end{align}
\end{subequations}
By convention, a polynomial with a negative degree must vanish.

We have studied the 1-point functions of the first 6 primary fields. Let us list these primary fields, and the combinatorial maps for the corresponding 10 solutions of modular covariance. For each solution, we indicate:
\begin{itemize}
\item which \underline{variables} the polynomials depend on,
 \item whether we need the \underline{sphere}-torus relation to derive the structure constants (red colour in the table indicates that we do),
 \item which sign factor $\epsilon \in \{+,-\}$ appears in the \underline{shift} equation $d_{(r,s+1)}= \epsilon d_{(r,s)}$ or $d_{(r,s+1)}= \epsilon d_{(r,s)}\big|_{w\to -w}$ (see discussion near Eq. \eqref{dddd}),
 \item whether the polynomials obey the \underline{parity} relation $d_{(r,s)}=d_{(r,-s)}$ (green colour in the table indicates that they do),
 \item how many nonvanishing \underline{poly}nomials we determined.
\end{itemize}
\begin{align}
\renewcommand{\arraystretch}{1.3}
 \begin{tabular}{|>{$}c<{$}|>{$}c<{$}|>{$}l<{$}|c|>{$}c<{$}|c|>{$}c<{$}|}
 \hline
 \text{Field} & \text{Map} & \text{Variables} & Sphere & \text{Shift} & \text{Parity} & \#\text{poly}
 \\
 \hline\hline
 V_{P_1} & (0, 0, 0) & n,w,w_1 & & + & \cellcolor{green!30} & 11
 \\
 \hline
 V_{(1,0)} & (1, 0, 0) & n,w &  & - & \cellcolor{green!30} & 10
 \\
 \hline
  \multirow{2}{*}{$V_{(2,0)}$} & (2, 0, 0) & n, w &   & + & \cellcolor{green!30} & 10
  \\ \cline{2-7} & (1, 1, 0) & n & & - & \cellcolor{green!30}  & 9
  \\
  \hline
  V_{(2,1)} & (2, 0, 0)  & n &  & + & \cellcolor{green!30} & 10
  \\
  \hline
  \multirow{3}{*}{$V_{(3,0)}$} & (3, 0, 0) & n,w &   & - &  \cellcolor{green!30} & 9
  \\ \cline{2-7} & (2,1,0) & n & \cellcolor{red!30} & - & \cellcolor{green!30} & 7
  \\ \cline{2-7} & (1, 1, 1) & n &  & + & \cellcolor{green!30} & 12
  \\
  \hline
  \multirow{2}{*}{$V_{(3,\frac23)}$} & (3, 0, 0) & n,w &   & - & & 6
  \\ \cline{2-7} & (2,1,0) & n & \cellcolor{red!30} & - &  & 7
  \\
  \hline
 \end{tabular}
 \label{tab}
\end{align}
The explicit results are found in Section \ref{sec:data}. We omit the odd spin fields $V_{(1, 1)}$, $V_{(2, \frac12)}$, $V_{(3, \frac13)}$ as their
torus one-point functions vanish, as we show below \eqref{som}.

\section{Combinatorial description of correlation functions}

In critical loop models, a non-diagonal primary field $V_{(r,s)}$ may be seen as a puncture inserting $2r$ legs.
Given a number of such fields on a Riemann surface, the various ways to connect the legs together can be encoded by a combinatorial map. To each combinatorial map, we can associate a correlation function in the CFT \cite{gjnrs23}.

We therefore expect that combinatorial maps parametrize a basis of solutions of conformal bootstrap equations. This expectation has been well-tested in the case of 4-point functions on the sphere, but not on Riemann surfaces of higher genus. We will now see what it implies for 1-point functions on the torus.

\subsection{One-point maps on the torus}
\label{subsec: maptor}

\subsubsection{Parametrization}

Consider a non-diagonal primary field $V_{(r_1,s_1)}$ on the torus.
For the correlation $\left<V_{(r_1,s_1)}\right>\neq 0$ to be non-zero,
the $2r_1$ legs must connect with one another, therefore $r_1\in\mathbb{N}$.
Let us enumerate the possible combinatorial maps for the first few values of $r_1$:
\begin{itemize}
 \item $\boxed{r_1=0}$ : there are no legs to connect, and only one trivial map.
 \item $\boxed{r_1=1}$ : the 2 legs must connect with one another, forming a loop that passes through the puncture. In a combinatorial map, two legs coming from the same puncture cannot be contracted together to form a contractible loop, thus the loop must be topologically nontrivial.
   Since combinatorial maps only account for the connectivities between the legs, all nontrivial topologies of the loop correspond to a single combinatorial map.
 \item $\boxed{r_1=2}$ : connecting 4 legs produces 2 loops, which we call $a$ and $b$. There are 2 distinct patterns for the cyclically ordered legs: $aabb$ and $abab$, corresponding to the loops being topologically the same or different. We therefore have 2 combinatorial maps.
 \item $\boxed{r_1=3}$ : the 3 loops can wrap 1, 2 or 3 topologically different cycles, leading to 3 combinatorial maps corresponding to the patterns $abccba$, $abcbac$ or $abcabc$.
\end{itemize}
For any $r_1\in\mathbb{N}$, the $r_1$ loops can wrap at most 3 topologically distinct cycles without intersecting. Therefore,
\begin{quote}
 combinatorial map $\simeq$ decomposition of $r_1$ as a sum of 3 integers, modulo cyclic permutations, or modulo all permutations if at least one integer vanishes.
\end{quote}
The set of maps is in bijection with:
\begin{align}
 \mathcal{M}(r_1) \simeq \left\{(m,n,p)\in\mathbb{N}^3\middle|m+n+p=r_1\middle|\begin{array}{l} m\geq n  \\ m\geq p \end{array} \middle| \begin{array}{l}
 n=0\implies p=0 \\ p=m\implies n=m \end{array}\right\}\ ,
\end{align}
where the conditions on $m,n,p$ account for cyclic reorderings of legs, and for the combinatorial equivalence of topologically different cycles. Let us draw the maps for $r_1=1, 2, 3$. We draw the torus as the square $\frac{\mathbb{C}}{\mathbb{Z}^2}$, with the puncture at the center:
\begin{align}
  \label{fig:cmaps-torus1p}
  \begin{tikzpicture}[baseline = (base)]
    \tore
      \draw (0, -1) to (0, 1);
      \node at (0, -1.4){(1, 0, 0)};
    \end{tikzpicture}
     \ \
  \begin{tikzpicture}[baseline = (base)]
    \tore
      \draw (0, -1) to (0, 1);
      \draw (0, 0) to (0.5, 1);
      \draw (0, 0) to (0.5, -1);
      \node at (0, -1.4){(2, 0, 0)};
    \end{tikzpicture}
     \ \
      \begin{tikzpicture}[baseline = (base)]
        \tore
        \draw (0, -1) to (0, 1);
        \draw (-1, 0) to (1, 0);
        \node at (0, -1.4){(1, 1, 0)};
      \end{tikzpicture}
    \ \
    \begin{tikzpicture}[baseline = (base)]
    \tore
      \draw (0, -1) to (0, 1);
      \draw (0, 0) to (-0.5, 1);
      \draw (0, 0) to (-0.5, -1);
      \draw (0, 0) to (0.5, 1);
      \draw (0, 0) to (0.5, -1);
      \node at (0, -1.4){(3, 0, 0)};
    \end{tikzpicture}
    \ \
      \begin{tikzpicture}[baseline = (base)]
      \tore
        \draw (0, -1) to (0, 1);
        \draw (-1, 0) to (1, 0);
        \draw (0, 0) to (0.5, 1);
        \draw (0, 0) to (0.5, -1);
        \node at (0, -1.4){(2, 1, 0)};
      \end{tikzpicture}
      \ \
      \begin{tikzpicture}[baseline = (base)]
        \tore
        \draw (0, -1) to (0, 1);
        \draw (-1, 0) to (1, 0);
        \draw (-1, -1) to (1, 1);
        \node at (0, -1.4){(1, 1, 1)};
      \end{tikzpicture}
\end{align}

\subsubsection{Symmetries of maps and constraints on the second Kac index}

The second Kac index of $V_{(r_1,s_1)}$ is an angular momentum around the puncture. It follows that symmetries of maps lead to constraints on $s_1$. In the case of a 2-point function on the sphere, $\left<V_{(r_1,s_1)}V_{(r_2,s_2)}\right>^\text{sphere}\neq 0 \implies (r_1,s_1)=(r_2,s_2)$, in agreement with conformal symmetry \cite{gjnrs23}. In the case of a 1-point function on the torus, if the map is invariant under a subgroup of cyclic permutations of the legs $\mathbb{Z}_q\subset \mathbb{Z}_{2r_1}$, then $\left<V_{(r_1,s_1)}\right>$ is invariant under a subgroup $\mathbb{Z}_q$ of rotations, so that
\begin{align}
 \left<V_{(r_1,s_1)}\right>\neq 0 \implies r_1s_1\in q\mathbb{Z} \ .
 \label{som}
\end{align}
The torus is invariant under the rotation by $\pi$ around the puncture, and
all maps have a $\mathbb{Z}_2$ symmetry generated by $r \in \mathbb{Z}_{2r}$, which implies $r_1s_1\in 2\mathbb{Z}$: the conformal spin must be even. Some particular maps have larger symmetries:
\begin{itemize}
 \item If $2|r_1$, the map $(\frac{r_1}{2},\frac{r_1}{2},0)$ has a $\mathbb{Z}_4$ symmetry.
 \item If $3|r_1$, the map $(\frac{r_1}{3},\frac{r_1}{3},\frac{r_1}{3})$ has a $\mathbb{Z}_6$ symmetry.
\end{itemize}
For $r_1\leq 8$, let us display all maps with larger symmetries, and the associated constraints on $s_1$:
\begin{equation}
\label{eq:94}
      \begin{tikzpicture}[baseline = (base)]
      \tore
        \draw (0, -1) to (0, 1);
        \draw (-1, 0) to (1, 0);
        \node at (0, -1.4){(1, 1, 0)};
        \node at (0, -2){$s_1\in 2\mathbb{Z}$};
      \end{tikzpicture}
   \ \
      \begin{tikzpicture}[baseline = (base)]
       \tore
        \draw (0, -1) to (0, 1);
        \draw (-1, 0) to (1, 0);
        \draw (-1, -1) to (1, 1);
        \node at (0, -1.4){(1, 1, 1)};
       \node at (0, -2){$s_1\in 2\mathbb{Z}$};
      \end{tikzpicture}
   \ \
    \begin{tikzpicture}[baseline = (base)]
    \tore
      \draw (0, 0) to (-0.2, 1);
      \draw (0, 0) to (-0.2, -1);
      \draw (0, 0) to (0.2, 1);
      \draw (0, 0) to (0.2, -1);
      \draw (0, 0) to (1, 0.2);
      \draw (0, 0) to (1, -0.2);
      \draw (0, 0) to (-1, 0.2);
      \draw (0, 0) to (-1, -0.2);
      \node at (0, -1.4){(2, 2, 0)};
     \node at (0, -2){$s_1\in \mathbb{Z}$};
    \end{tikzpicture}
      \ \
    \begin{tikzpicture}[baseline = (base)]
      \tore
      \draw (0, 0) to (0, -1);
      \draw (0, 0) to (0, 1);
      \draw (0, 0) to (-0.15, 1);
      \draw (0, 0) to (-0.15, -1);
      \draw (0, 0) to (0.15, 1);
      \draw (0, 0) to (0.15, -1);
      \draw (0, 0) to (1, 0.15);
      \draw (0, 0) to (1, -0.15);
      \draw (0, 0) to (-1, 0.15);
      \draw (0, 0) to (-1, -0.15);
      \draw (0, 0) to (1, 0);
      \draw (0, 0) to (-1, 0);
      \node at (0, -1.4){(3, 3, 0)};
     \node at (0, -2){$s_1\in \frac23\mathbb{Z}$};
    \end{tikzpicture}
   \ \
    \begin{tikzpicture}[baseline = (base)]
   \tore
      \draw (0, 0) to (-0.2, 1);
      \draw (0, 0) to (-0.2, -1);
      \draw (0, 0) to (0.2, 1);
      \draw (0, 0) to (0.2, -1);
      \draw (0, 0) to (1, 0.2);
      \draw (0, 0) to (1, -0.2);
      \draw (0, 0) to (-1, 0.2);
      \draw (0, 0) to (-1, -0.2);
      \draw (-1, -0.8) to (1, 0.8);
      \draw (-0.8, -1) to (0.8, 1);
      \draw (-1, .8) to (-.8, 1);
      \draw (1, -.8) to (.8, -1);
      \node at (0, -1.4){(2, 2, 2)};
     \node at (0, -2){$s_1\in \mathbb{Z}$};
    \end{tikzpicture}
   \ \
    \begin{tikzpicture}[baseline = (base)]
    \tore
      \draw (0, 0) to (-0.125, 1);
      \draw (0, 0) to (-0.125, -1);
      \draw (0, 0) to (0.125, 1);
      \draw (0, 0) to (0.125, -1);
      \draw (0, 0) to (-0.35, 1);
      \draw (0, 0) to (-0.35, -1);
      \draw (0, 0) to (0.35, 1);
      \draw (0, 0) to (0.35, -1);
      \draw (0, 0) to (1, 0.125);
      \draw (0, 0) to (1, -0.125);
      \draw (0, 0) to (-1, 0.125);
      \draw (0, 0) to (-1, -0.125);
      \draw (0, 0) to (1, 0.35);
      \draw (0, 0) to (1, -0.35);
      \draw (0, 0) to (-1, 0.35);
      \draw (0, 0) to (-1, -0.35);
      \node at (0, -1.4){(4, 4, 0)};
     \node at (0, -2){$s_1\in \frac12\mathbb{Z}$};
    \end{tikzpicture}
\end{equation}
Taking into account these symmetries, the number of maps that lead to nonzero 1-point functions $\left<V_{(r_1,s_1)}\right>$ is therefore
\begin{align}
 \mathcal{N}_{(r_1,s_1)} = |\mathcal{M} (r_1)| - \delta_{2|r_1}(1-\delta_{4|r_1s_1}) - \delta_{3|r_1}(1-\delta_{6|r_1s_1})\ .
 \label{nroso}
\end{align}
Depending on $s_1$, we may therefore lose 0, 1 or 2 maps from $\mathcal{M} (r_1)$.

\subsubsection{Parity}

If $m>n>p>0$, the decomposition $r_1=m+n+p$ gives rise to two different maps $(m,n,p)$ and $(m,p,n)$. These two maps are related by a parity transformation $z\to \bar z$. For a primary field that is parity-invariant i.e. $s_1=0$, these two maps can be linearly combined into two correlation functions of the type $\left<V_{(r_1,0)}\right>$:
a parity-even function such that $\left<V_{(r_1,0)}(z_1)\right>=\left<V_{(r_1,0)}(\bar z_1)\right>$, and a parity-odd function such that $\left<V_{(r_1,0)}(z_1)\right>=-\left<V_{(r_1,0)}(\bar z_1)\right>$. On the other hand, a correlation function $\left<V_{(r_1,0)}(z_1)\right>$ corresponding to a parity-invariant combinatorial map is parity-even.
The set of parity-invariant maps is
\begin{align}
 \mathcal{M}^0(r_1) = \left\{(m,n,p)\in\mathbb{N}^3\middle|m+n+p=r_1\middle|
 \begin{array}{r}
 m\geq n\geq p\geq 0\\ \text{not } m>n>p>0
 \end{array}\right\}\ .
\end{align}
We also define two sets of maps that are in bijection with sets of parity-even and parity-odd correlation functions respectively:
\begin{align}
 \mathcal{M}^+(r_1) &= \left\{(m,n,p)\in\mathbb{N}^3\middle|m+n+p=r_1\middle|
 m\geq n\geq p\geq 0\right\}\ ,
 \\
 \mathcal{M}^-(r_1) &= \left\{(m,n,p)\in\mathbb{N}^3\middle|m+n+p=r_1\middle|
 m>n>p>0\right\}\ ,
\end{align}
so that
\begin{align}
 \mathcal{M}^+(r_1) = \mathcal{M}^-(r_1) \sqcup \mathcal{M}^0(r_1) \quad , \quad \mathcal{M}(r_1)\simeq \mathcal{M}^+(r_1) \sqcup \mathcal{M}^-(r_1)\ .
\end{align}
(The second one of these bijections can be realized using the injection $(m,n,p)\mapsto (m,p,n)$ from $\mathcal{M}^-(r_1)$ to $\mathcal{M}(r_1)$.)

\subsubsection{Counting maps}

In order to determine numbers of maps, let us start from the elementary identities:
\begin{align}
 |\mathcal{M}^0(r_1)|=r_1 \quad ,\quad |\mathcal{M}^+(r_1)|=|\mathcal{M}^-(r_1+6)|\ .
\end{align}
The first identity is proved by induction on $r_1$:
\begin{quote}
Let $\psi:\mathcal{M}^0(r_1) \to \mathcal{M}^0(r_1+1)$ be defined by $\psi((m,n,p))=(n,m,p+1)$ if $m=n> p > 0$, otherwise $\psi((m,n,p))=(m+1,n,p)$. Then $\psi$ is injective, and $\mathcal{M}^0(r_1+1)=\psi(\mathcal{M}^0(r_1))\sqcup \{(\lfloor\frac{r_1}{2}\rfloor,\lfloor\frac{r_1}{2}\rfloor,\delta_{r_1\in 2\mathbb{N}})\}$.
\end{quote}
The second identity follows from the bijection $(m,n,p)\mapsto (m+3,n+2,p+1)$. We deduce $|\mathcal{M}^-(r_1+6)| = |\mathcal{M}^-(r_1)|+r_1$, thus we can determine $|\mathcal{M}^-(r_1)|$ from its first 6 values, namely $|\mathcal{M}^-(r_1\leq 5)|=0$ and $|\mathcal{M}^-(6)|=1$. This leads to the results
\begin{align}
 |\mathcal{M}^\pm (r_1)| = \left\lfloor \frac{{(r_1\pm 3)}^2}{12} \right\rceil \quad , \quad |\mathcal{M}(r_1)| = \left\lfloor \frac{r_1^2+9}{6}\right\rfloor + \delta_{r_1 \equiv 0 \bmod 6}-\delta_{r_1=0} \ ,
 \label{mr1}
\end{align}
where $\lfloor x \rceil$ is the nearest integer to $x\in \mathbb{R}\backslash (\mathbb{Z}+\frac12)$, while $\lfloor x\rfloor$ is the floor. The corresponding generating functions are
\begin{align}
 \sum_{r_1=0}^\infty |\mathcal{M} (r_1)|\, x^{r_1} &= \frac{1+x^6}{(1-x)(1-x^2)(1-x^3)}\ ,
 \\
 \sum_{r_1=0}^\infty |\mathcal{M}^+ (r_1)|\, x^{r_1} &= \frac{1}{(1-x)(1-x^2)(1-x^3)} \ .
\end{align}
The first few values are:
\begin{align}
  \label{tab:Mc11}
  \renewcommand{\arraystretch}{1.5}
\begin{array}{|r | c c c c c c c c c c c c c c c|}
\hline
r_1 & 0 & 1 & 2 & 3 & 4 & 5 & 6 & 7 & 8 & 9 & 10 & 11 & 12 & 13 & 14\\ \hline\hline
|\mathcal M(r_1)| & 1 & 1 & 2 & 3 & 4 & 5 & 8 & 9 & 12 & 15 & 18 & 21 & 26 & 29 & 34\\ \hline
|\mathcal M^-(r_1)| & 0 & 0 & 0 & 0 & 0 & 0 & 1 & 1 & 2 & 3 & 4 & 5 & 7 & 8 & 10
\\ \hline
|\mathcal M^+(r_1)| & 1 & 1 & 2 & 3 & 4 & 5 & 7 & 8 & 10 & 12 & 14 & 16 & 19 & 21 & 24
\\
\hline
\end{array}
\end{align}

\subsection{Relation with four-point maps on the sphere}

\subsubsection{Why do we need the sphere?}

Combinatorial maps allow us to predict dimensions of spaces of solutions of conformal bootstrap equations. However, we also want to determine the particular solution associated to any given map. On the sphere, maps are known to provide a basis of solutions that have tractable analytic expressions \cite{nrj23}, and the results of section \ref{sec:data} show that the same holds on the torus. Our goal is thus to associate a solution of bootstrap equations to any given combinatorial map.

There are three combinatorially distinct types of loops on the torus with one puncture:
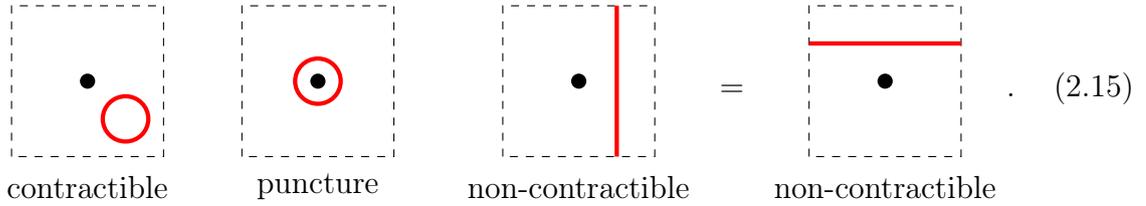
\begin{align}
  \label{fig:loops}
 \begin{tikzpicture}[baseline = (base)]
      \tore
        \draw[ultra thick, red] (.5, -.5) circle (.3);
        \node at (0, -1.4){contractible};
      \end{tikzpicture}
   \qquad
      \begin{tikzpicture}[baseline = (base)]
       \tore
        \draw[ultra thick, red] (0, 0) circle (.3);
        \node at (0, -1.4){puncture};
      \end{tikzpicture}
   \qquad
    \begin{tikzpicture}[baseline = (base)]
    \tore
      \draw[ultra thick, red] (.5, -1) to (.5, 1);
      \node at (0, -1.4){non-contractible};
    \end{tikzpicture}
    \ =  \
     \begin{tikzpicture}[baseline = (base)]
    \tore
      \draw[ultra thick, red] (-1, .5) to (1, .5);
      \node at (0, -1.4){non-contractible};
    \end{tikzpicture}.
\end{align}
Contractible loops and puncture loops are not relevant to bootstrap equations, nor do they capture any information on our combinatorial map. We are interested in non-contractible loops. All of the topologically distinct non-contractible loops are combinatorially equivalent, since slicing the torus along any non-contractible loop results in a cylinder.

We define the signature $\sigma\in\frac12\mathbb{N}$ of a map as half the minimum number of intersections of the map with a non-contractible loop. (We care about the signature because it has a simple interpretation in terms of the corresponding correlation function.)
The loop around the cycle with the largest number of wrapped legs minimizes the number of intersections:
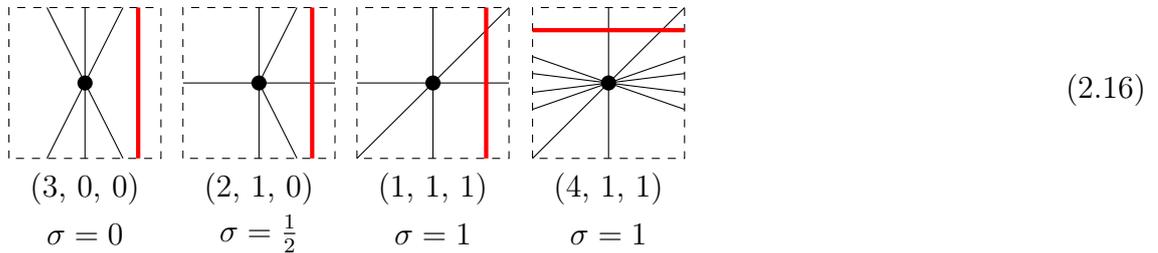
\begin{align}
  \label{fig:sigs}
    \begin{tikzpicture}[baseline = (base)]
    \tore
      \draw (0, -1) to (0, 1);
      \draw (0, 0) to (-0.5, 1);
      \draw (0, 0) to (-0.5, -1);
      \draw (0, 0) to (0.5, 1);
      \draw (0, 0) to (0.5, -1);
      \draw[ultra thick, red] (.7, -1) to (.7, 1);
      \node at (0, -1.4){(3, 0, 0)};
      \node at (0, -2){$\sigma=0$};
    \end{tikzpicture}
    \ \
      \begin{tikzpicture}[baseline = (base)]
      \tore
        \draw (0, -1) to (0, 1);
        \draw (-1, 0) to (1, 0);
        \draw (0, 0) to (0.5, 1);
        \draw (0, 0) to (0.5, -1);
        \node at (0, -1.4){(2, 1, 0)};
         \draw[ultra thick, red] (.7, -1) to (.7, 1);
         \node at (0, -2){$\sigma=\frac12$};
      \end{tikzpicture}
      \ \
      \begin{tikzpicture}[baseline = (base)]
        \tore
        \draw (0, -1) to (0, 1);
        \draw (-1, 0) to (1, 0);
        \draw (-1, -1) to (1, 1);
        \node at (0, -1.4){(1, 1, 1)};
         \draw[ultra thick, red] (.7, -1) to (.7, 1);
         \node at (0, -2){$\sigma=1$};
      \end{tikzpicture}
       \ \
     \begin{tikzpicture}[baseline = (base)]
    \tore
      \draw (0, -1) to (0, 1);
      \draw (-1, -1) to (1, 1);
      \draw (0, 0) to (1, 0.125);
      \draw (0, 0) to (1, -0.125);
      \draw (0, 0) to (-1, 0.125);
      \draw (0, 0) to (-1, -0.125);
      \draw (0, 0) to (1, 0.35);
      \draw (0, 0) to (1, -0.35);
      \draw (0, 0) to (-1, 0.35);
      \draw (0, 0) to (-1, -0.35);
      \draw[ultra thick, red] (-1, .7) to (1, .7);
      \node at (0, -1.4){(4, 1, 1)};
     \node at (0, -2){$\sigma=1$};
    \end{tikzpicture}
\end{align}
The signature of an arbitrary map is therefore
\begin{align}
 \sigma(m,n,p) = \frac12(r_1-m) = \frac12(n+p) \ .
 \label{sig}
\end{align}
Hence the signature does not characterize the map $(m,n,p)$.
To be precise, a map $M$ is characterized by its signature if $M'\neq M\implies \sigma(M')<\sigma(M)$ \cite{gjnrs23}.
On the torus, only the following maps are characterized by their signatures:
\begin{align}
 (1,0, 0) \quad ,\quad \Big\{(m,m,m)\Big\}_{m\in\mathbb{N}} \quad, \quad \Big\{(m,m,m-1)\Big\}_{m\in\mathbb{N}^*}\ .
 \label{mcbs}
\end{align}
In general, several maps can have the same signature, and we do not know how to distinguish the corresponding solutions of bootstrap equations. The situation would be much better if we could distinguish the 3 different cycles that are wrapped by the legs. This is what the sphere with 4 punctures allows us to do.

\subsubsection{From the torus to the sphere}

Let us justify the introduction of a sphere with 4 punctures, and derive the nature of these punctures.
\begin{itemize}
 \item On the torus, a puncture with $2r_1$ legs gives rise to $r_1$ loops that wrap at most 3 topologically different cycles. In order to have a counterpart of this situation on the sphere, we consider a sphere with 3 legless punctures, in addition to the puncture with $2r_1$ legs. The $r_1$ loops can then wrap 3 topologically different cycles without intersecting.
 \item Now consider closed loops, i.e. loops that do not go through the puncture. A non-contractible closed loop on the torus corresponds to a closed loop that separates the sphere punctures as $4=2+2$.
 \item No loop on the torus maps to a closed loop around one of the 3 legless punctures on the sphere. Therefore, such a closed loop on the sphere must have weight $w=0$.
\end{itemize}
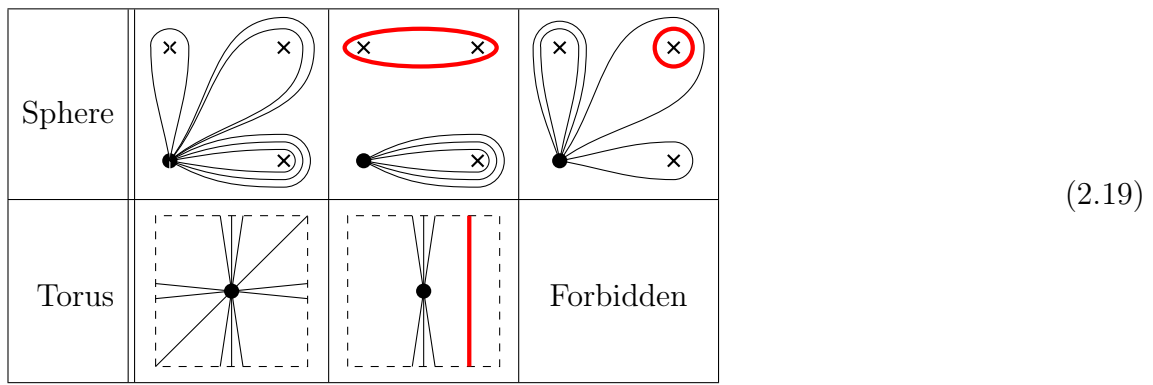
\begin{align}
 \begin{array}{|r||c|c|c|}
  \hline
  \text{Sphere} &
\begin{tikzpicture}[baseline = (base), scale = .5]
 \vertices
 \draw[white] (0, -1) -- (0, 4);
 \draw (0, 0) to [out=80, in=-90] (0.5, 3)  to [out=90, in=0] (0, 3.5) to [out=180, in=100] (-0.5, 3) to [out=-90, in=100] (0, 0);
  \draw (0, 0) to [out=-10, in=-180] (3, -0.3)  to [out=0, in=-90] (3.3, 0) to [out=90, in=0] (3, 0.3) to [out=180, in=10] (0, 0);
  \draw (0, 0) to [out=-20, in=-180] (3, -0.5)  to [out=0, in=-90] (3.5, 0) to [out=90, in=0] (3, 0.5) to [out=180, in=20] (0, 0);
  \draw (0, 0) to [out=-30, in=-180] (3, -0.7)  to [out=0, in=-90] (3.7, 0) to [out=90, in=0] (3, 0.7) to [out=180, in=30] (0, 0);
   \draw (0, 0) to [out=40, in=-90] (3.5, 3) to [out=90, in=0] (3, 3.5) to [out=180, in=50] (0, 0);
          \draw (0, 0) to [out=37, in=-90] (3.8, 3) to [out=90, in=0] (3, 3.8) to [out=180, in=53] (0, 0);
\end{tikzpicture}
  &
  \begin{tikzpicture}[baseline = (base), scale = .5]
 \vertices
  \draw (0, 0) to [out=-10, in=-180] (3, -0.3)  to [out=0, in=-90] (3.3, 0) to [out=90, in=0] (3, 0.3) to [out=180, in=10] (0, 0);
  \draw (0, 0) to [out=-20, in=-180] (3, -0.5)  to [out=0, in=-90] (3.5, 0) to [out=90, in=0] (3, 0.5) to [out=180, in=20] (0, 0);
  \draw (0, 0) to [out=-30, in=-180] (3, -0.7)  to [out=0, in=-90] (3.7, 0) to [out=90, in=0] (3, 0.7) to [out=180, in=30] (0, 0);
          \draw[ultra thick, red] plot [smooth cycle, tension = 1] coordinates {(-.5,3) (1.5,2.5) (3.5,3) (1.5,3.5)};
\end{tikzpicture}
  &
  \begin{tikzpicture}[baseline = (base), scale = .5]
 \vertices
 \draw (0, 0) to [out=80, in=-90] (0.5, 3)  to [out=90, in=0] (0, 3.5) to [out=180, in=100] (-0.5, 3) to [out=-90, in=100] (0, 0);
          \draw (0, 0) to [out=70, in=-90] (0.7, 3)  to [out=90, in=0] (0, 3.7) to [out=180, in=90] (-0.7, 3) to [out=-90, in=110] (0, 0);
  \draw (0, 0) to [out=-10, in=-180] (3, -0.5)  to [out=0, in=-90] (3.5, 0) to [out=90, in=0] (3, 0.5) to [out=180, in=10] (0, 0);
          \draw (0, 0) to [out=37, in=-90] (3.8, 3) to [out=90, in=0] (3, 3.8) to [out=180, in=53] (0, 0);
          \draw [ultra thick, red] (3, 3) circle (.5);
\end{tikzpicture}
  \\ \hline
  \text{Torus} &
   \begin{tikzpicture}[baseline = (base)]
      \draw[white] (0, -1.2) -- (0, 1.2);
      \tore
      \draw (0, 0) to (0, -1);
      \draw (0, 0) to (0, 1);
      \draw (0, 0) to (-0.15, 1);
      \draw (0, 0) to (-0.15, -1);
      \draw (0, 0) to (0.15, 1);
      \draw (0, 0) to (0.15, -1);
      \draw (0, 0) to (1, 0.1);
      \draw (0, 0) to (1, -0.1);
      \draw (0, 0) to (-1, 0.1);
      \draw (0, 0) to (-1, -0.1);
      \draw (-1, -1) to (1, 1);
    \end{tikzpicture}
    &
     \begin{tikzpicture}[baseline = (base)]
    \tore
      \draw (0, -1) to (0, 1);
      \draw (0, 0) to (-0.15, 1);
      \draw (0, 0) to (-0.15, -1);
      \draw (0, 0) to (0.15, 1);
      \draw (0, 0) to (0.15, -1);
      \draw[ultra thick, red] (.6, -1) to (.6, 1);
    \end{tikzpicture}
    &
    \text{Forbidden}
    \\ \hline
 \end{array}
 \label{stmaps}
\end{align}
It follows that our sphere 4-point function must be
$\left<V'_0V_{(r_1,\frac{s_1}{2})}V'_0V'_0\right>^\text{sphere}$, where the field $V_0$ was
defined after Eq. \eqref{wP}. The value of the second Kac index $\frac{s_1}{2}$ is such that
the conformal spin of $V_{(r_1,\frac{s_1}{2})}$ is half that of $V_{(r_1,s_1)}$, as expected because all maps and correlation functions are invariant under rotations by $2\pi$ on the sphere versus $\pi$ on the torus.

The 3 extra punctures are labelled, and give rise to 3 cycles that we can distinguish. A combinatorial map is therefore characterized by the numbers of loops $m,n,p$ around each puncture, without allowing for permutations of the punctures. We still have to distinguish two cases:
\begin{itemize}
 \item $\boxed{mnp=0}$ : the numbers $m,n,p$ characterize a single map, which we call $(m,n,p)^*$, in order to distinguish it from the torus map $(m,n,p)$.

 \item $\boxed{mnp\neq 0}$ : the numbers $m,n,p$ characterize two maps $(m,n,p)^\pm$, which differ by the cyclic ordering of the loops around $V_{(r_1,s_1)}$, and are therefore related by a parity transformation. For example:
\begin{align}
\begin{tikzpicture}[baseline = (base), scale = .5]
 \vertices
 \draw (0, 0) to [out=80, in=-90] (0.5, 3)  to [out=90, in=0] (0, 3.5) to [out=180, in=100] (-0.5, 3) to [out=-90, in=100] (0, 0);
          \draw (0, 0) to [out=70, in=-90] (0.7, 3)  to [out=90, in=0] (0, 3.7) to [out=180, in=90] (-0.7, 3) to [out=-90, in=110] (0, 0);
 \node at (1.5, -1.4){$(1,2,2)^+$};
  \draw (0, 0) to [out=-10, in=-180] (3, -0.5)  to [out=0, in=-90] (3.5, 0) to [out=90, in=0] (3, 0.5) to [out=180, in=10] (0, 0);
   \draw (0, 0) to [out=40, in=-90] (3.5, 3) to [out=90, in=0] (3, 3.5) to [out=180, in=50] (0, 0);
          \draw (0, 0) to [out=37, in=-90] (3.8, 3) to [out=90, in=0] (3, 3.8) to [out=180, in=53] (0, 0);
\end{tikzpicture}
\qquad
\begin{tikzpicture}[baseline = (base), scale = .5]
 \vertices
 \draw (0, 0) to [out=80, in=-90] (0.5, 3)  to [out=90, in=0] (0, 3.5) to [out=180, in=100] (-0.5, 3) to [out=-90, in=100] (0, 0);
          \draw (0, 0) to [out=70, in=-90] (0.7, 3)  to [out=90, in=0] (0, 3.7) to [out=180, in=90] (-0.7, 3) to [out=-90, in=110] (0, 0);
 \node at (1.5, -1.4){$(1,2,2)^-$};
  \draw (0, 0) to [out=-10, in=-180] (3, -0.5)  to [out=0, in=-90] (3.5, 0) to [out=90, in=0] (3, 0.5) to [out=180, in=10] (0, 0);
  \draw (0, 0) to [out=150, in=180] (0, 4.1) to [out=0, in=-180] (3, 2.4) to [out=0, in=-90](3.5, 3) to [out=90, in=0] (0, 4.7) to [out=180, in=90] (-1.7, 1.5) to [out=-90,in=-150](0, 0);
  \draw (0, 0) to [out=170, in=180] (0, 4.3) to [out=0, in=-90] (3.3, 3) to [out=90, in=0] (0, 4.5) to [out=180, in=90] (-1.4, 1.5) to [out=-90,in=-170](0, 0);
\end{tikzpicture}
\end{align}
\end{itemize}
Mapping the sphere to the torus amounts to forgetting the labels of the 3 extra punctures, while remembering the cyclic ordering of legs around the original puncture. For a given map $M\in \mathcal{M}(r_1)$ on the torus, let $\varphi(M)\subset \mathcal{M}^\text{sphere}(r_1)$ be the set of corresponding maps on the sphere, then $|\varphi(M)|\in\{1,2,3,6\}$. Let us indicate the different cases, using integers $m>n>p>0$:
\begin{subequations}\label{phi}
\begin{align}
 \varphi(m,n,p) &= \Big\{(m,n,p)^+, (m,p,n)^-, (n,p,m)^+, (n,m,p)^-, (p,m,n)^+, (p,n,m)^-\Big\}\ ,
 \\
 \varphi(m,m,p) &= \Big\{(m,m,p)^\pm,(m,p,m)^\pm , (p,m,m)^\pm\Big\}\ ,
 \\
 \varphi(m,n,n) &= \Big\{(m,n,n)^\pm, (n,m,n)^\pm,(n,n,m)^\pm\Big\}\ ,
 \\
 \varphi(m,m,m) &= \Big\{(m,m,m)^\pm \Big\}\ ,
 \\
 \varphi(m,n,0) &= \Big\{(m,n,0)^*,(n,m,0)^*,(m,0,n)^*,(n,0,m)^*,(0,m,n)^*,(0,n,m)^*\Big\}\ ,
 \\
 \varphi(m,m,0) &= \Big\{(m,m,0)^*, (m,0,m)^*, (0,m,m)^*\Big\}\ ,
 \\
 \varphi(m,0,0) &= \Big\{(m,0,0)^*,(0,m,0)^*,(0,0,m)^*\Big\}\ ,
 \\
 \varphi(0,0,0) &= \Big\{(0,0,0)^*\Big\}\ .
\end{align}
\end{subequations}
The number of 4-point maps on the sphere is known to be $|\mathcal{M}^\text{sphere}(r_1)|=r_1^2+2-\delta_{r_1=0}$ \cite{gjnrs23}, to be compared with $|\mathcal{M}(r_1)|$ \eqref{mr1}. We have $\mathcal{M}^\text{sphere}(r_1) =\sqcup_{M\in \mathcal{M}(r_1)}\varphi(M)$, consistently with $|\varphi(M)|=6$ for generic maps.

\subsubsection{Signatures on the sphere}

For a 4-point map on the sphere, the signature is a triple $\sigma^*\in (\frac12\mathbb{N})^3$, which counts the intersections of the map with the 3 loops that separate our punctures into pairs. For maps with 3 legless punctures, these half-integers are in fact integers. Moreover, the signature of $(m,n,p)^\pm$ does not depend on the sign $\pm$, and is
\begin{align}
 \sigma^*(m,n,p) =(\sigma^{(s)},\sigma^{(t)},\sigma^{(u)}) = (m+n,n+p,m+p)\ .
\end{align}
By definition the torus signature is $\sigma(m,n,p) = \frac12\min\sigma^*(m,n,p)$, which agrees with Eq. \eqref{sig}. For example, in the case $(m,n,p)=(3,2,1)$, the torus and sphere signatures are:
\begin{align}
  \begin{tikzpicture}[baseline = (base)]
    \tore
     \draw[ultra thick, red] (-1, .6) to (1, .6);
    \draw (0, -1) to (0, 1);
     \draw (-1, -0.8) to (1, 0.8);
      \draw (-0.8, -1) to (0.8, 1);
      \draw (-1, .8) to (-.8, 1);
      \draw (1, -.8) to (.8, -1);
      \draw (-1, 0) to (1, 0);
      \draw (-1, .2) to (1, -.2);
      \draw (-1, -.2) to (1, .2);
      \node at (0, -1.4){(3, 2, 1)};
     \node at (0, -2){$\sigma = \frac32$};
    \end{tikzpicture}
    \hspace{2cm}
    \begin{tikzpicture}[baseline = (base), scale = .5]
 \vertices
 \draw (0, 0) to [out=80, in=-90] (0.5, 3)  to [out=90, in=0] (0, 3.5) to [out=180, in=100] (-0.5, 3) to [out=-90, in=100] (0, 0);
  \draw (0, 0) to [out=-10, in=-180] (3, -0.3)  to [out=0, in=-90] (3.3, 0) to [out=90, in=0] (3, 0.3) to [out=180, in=10] (0, 0);
  \draw (0, 0) to [out=-20, in=-180] (3, -0.5)  to [out=0, in=-90] (3.5, 0) to [out=90, in=0] (3, 0.5) to [out=180, in=20] (0, 0);
  \draw (0, 0) to [out=-30, in=-180] (3, -0.7)  to [out=0, in=-90] (3.7, 0) to [out=90, in=0] (3, 0.7) to [out=180, in=30] (0, 0);
   \draw (0, 0) to [out=40, in=-90] (3.5, 3) to [out=90, in=0] (3, 3.5) to [out=180, in=50] (0, 0);
          \draw (0, 0) to [out=37, in=-90] (3.8, 3) to [out=90, in=0] (3, 3.8) to [out=180, in=53] (0, 0);
          \draw[ultra thick, red] plot [smooth cycle, tension = 1.5] coordinates {(0,-.7) (1,1.5) (0,3.7) (-1,1.5)};
          \node at (1.5, -1.4){$(3,2,1)^+$};
          \node at (1.5, -2.6){$\sigma^{(s)} = 5$};
\end{tikzpicture}
\quad
\begin{tikzpicture}[baseline = (base), scale = .5]
 \vertices
 \draw (0, 0) to [out=80, in=-90] (0.5, 3)  to [out=90, in=0] (0, 3.5) to [out=180, in=100] (-0.5, 3) to [out=-90, in=100] (0, 0);
  \draw (0, 0) to [out=-10, in=-180] (3, -0.3)  to [out=0, in=-90] (3.3, 0) to [out=90, in=0] (3, 0.3) to [out=180, in=10] (0, 0);
  \draw (0, 0) to [out=-20, in=-180] (3, -0.5)  to [out=0, in=-90] (3.5, 0) to [out=90, in=0] (3, 0.5) to [out=180, in=20] (0, 0);
  \draw (0, 0) to [out=-30, in=-180] (3, -0.7)  to [out=0, in=-90] (3.7, 0) to [out=90, in=0] (3, 0.7) to [out=180, in=30] (0, 0);
   \draw (0, 0) to [out=40, in=-90] (3.5, 3) to [out=90, in=0] (3, 3.5) to [out=180, in=50] (0, 0);
          \draw (0, 0) to [out=37, in=-90] (3.8, 3) to [out=90, in=0] (3, 3.8) to [out=180, in=53] (0, 0);
          \draw[ultra thick, red] plot [smooth cycle, tension = 1.5] coordinates {(-1,3) (1.5,2) (4,3) (1.5,4.2)};
          \node at (1.5, -1.4){$(3,2,1)^+$};
          \node at (1.5, -2.6){$\sigma^{(t)} = 3$};
\end{tikzpicture}
\quad
\begin{tikzpicture}[baseline = (base), scale = .5]
 \vertices
 \draw (0, 0) to [out=80, in=-90] (0.5, 3)  to [out=90, in=0] (0, 3.5) to [out=180, in=100] (-0.5, 3) to [out=-90, in=100] (0, 0);
  \draw (0, 0) to [out=-10, in=-180] (3, -0.3)  to [out=0, in=-90] (3.3, 0) to [out=90, in=0] (3, 0.3) to [out=180, in=10] (0, 0);
  \draw (0, 0) to [out=-20, in=-180] (3, -0.5)  to [out=0, in=-90] (3.5, 0) to [out=90, in=0] (3, 0.5) to [out=180, in=20] (0, 0);
  \draw (0, 0) to [out=-30, in=-180] (3, -0.7)  to [out=0, in=-90] (3.7, 0) to [out=90, in=0] (3, 0.7) to [out=180, in=30] (0, 0);
   \draw (0, 0) to [out=40, in=-90] (3.5, 3) to [out=90, in=0] (3, 3.5) to [out=180, in=50] (0, 0);
          \draw (0, 0) to [out=37, in=-90] (3.8, 3) to [out=90, in=0] (3, 3.8) to [out=180, in=53] (0, 0);
          \draw[ultra thick, red] plot [smooth cycle, tension = 1.5] coordinates {(-.5,-.5) (4,2) (2, 4)};
          \node at (1.5, -1.4){$(3,2,1)^+$};
          \node at (1.5, -2.6){$\sigma^{(u)} = 4$};
\end{tikzpicture}
\end{align}
The sphere signature contains more information than the torus signature: is this enough for characterizing the corresponding maps? For our particular 4-point maps, we have
\begin{align}
 & \sigma^*(M')\geq \sigma^*((m,n,p)^*)\implies M'=(m,n,p)^*\ ,
 \\
 & \sigma^*(M')\geq \sigma^*((m,n,p)^\epsilon)\implies M' \in \Big\{(m,n,p)^+,(m,n,p)^-\Big\}\ .
\end{align}
for any $\epsilon \in\{+,-\}$. Therefore, maps with $mnp\neq 0$ are characterized by their signatures up to a twofold parity ambiguity.

\subsection{Solutions of conformal bootstrap equations}\label{sec:scbe}

Given a 1-point map $M$ on the torus, we call $Z_M$ the corresponding 1-point function. Similarly, we call $Z^*_{M'}$ the 4-point function associated to a 4-point map $M'$ on the sphere. We conjecture that these solutions are related by
\begin{align}
 Z_M' = \sum_{M'\in \varphi(M)} Z^*_{M'}\ .
 \label{pzz}
\end{align}
This means that after performing the change of variables $Z_M \to Z'_M$ according to \eqref{rel}, $Z_M$ becomes the sum of the sphere 4-point functions for all sphere maps that correspond to $M$. We will now see what this relation implies, and how it helps us compute $Z_M$.

\subsubsection{Torus and sphere spectrums}

Conformal bootstrap equations arise from decompositions of correlation functions into conformal blocks. The blocks that appear in a given decomposition are determined by the corresponding spectrum of primary fields. These spectrums are constrained by fusion rules. In critical loop models, fusion rules say that the first Kac index is conserved modulo integers. Moreover, given a combinatorial map, the first Kac index is bounded from below by the map's signature.
For a torus 1-point function $\left<V_{(r_1,s_1)}\right>$, the channel spectrum is the full spectrum $\mathcal{S}$ \eqref{spec}, truncated according to the signature:
\begin{align}
 \mathcal{S}_\sigma=
 \left\{V_{(r,s)}\right\}_{\substack{r\in \sigma+\frac12\mathbb{N}\\ s\in\frac{1}{r}\mathbb{Z}}}\, , \quad
 \begin{tikzpicture}[baseline = (base), scale = .6]
   \coordinate (base) at (0, 0);
   \node[below] at (-2.3, 0) {$V_{(r_1,s_1)}$};
   \node[right] at (3, 0) {$V_{(r, s)}$};
   \draw[ultra thick] (-2.5, 0) to (0, 0);
  \draw[ultra thick] plot [smooth, tension = .8] coordinates {(0, 0) (2, 1) (3, 0) (2, -1) (0, 0)};
 \end{tikzpicture}
\end{align}
For a sphere 4-point function $\left<V_0V_1V_0V_0\right>^\text{sphere}$, given a channel $x\in\{s,t,u\}$, the $x$-channel spectrum is truncated according to the signature $\sigma^{(x)}$, and the first Kac index must be integer. This leads to a spectrum $\mathcal{S}^{(x)}= \mathcal{S}^*_{\sigma^{(x)}}$, where we define
\begin{align}
 \mathcal{S}^*_{\sigma} =
 \left\{V_{(r,s)}\right\}_{\substack{r\in \sigma+\mathbb{N}\\ s\in\frac{1}{r}\mathbb{Z}}}\, , \quad
 \begin{tikzpicture}[baseline = (base), scale = .4]
   \coordinate (base) at (0, 0);
  \draw[ultra thick] (-1 ,2) -- (0, 0) -- (4, 0) -- (5, 2);
  \draw[ultra thick] (-1, -2) -- (0, 0);
  \draw[ultra thick] (5, -2) -- (4, 0);
  \node[left] at (-1, -2) {$V_0$};
  \node[right] at (5,2) {$V_0$};
  \node[right] at (5,-2) {$V_0$};
  \node[left] at (-.8,2) {$V_{(r_1,s_1)}$};
  \node[below] at (2, 0) {$V_{(r,s)}$};
 \end{tikzpicture}
\end{align}
From the torus to the sphere, the signature of a map, and therefore also the number of legs of a channel field, are multiplied by $2$. This explains why the sphere-torus relation \eqref{rel} is $V_{(r,s)}'= V_{(2r,s)}$ for channel fields. The sphere-torus relation therefore leads to sphere channel spectrums of the type
\begin{align}
 \mathcal{S}_\sigma' = \left. \mathcal{S}^*_{2\sigma}\right|_{rs\in 2\mathbb{Z}} \ .
\end{align}
Therefore, a sphere 4-point function that corresponds to a torus 1-point function has channel spectrums made of even spin fields.

\subsubsection{Determination of torus 1-point functions}

For a torus 1-point map $M=(m,n,p)$, let us discuss how we can compute the corresponding 1-point function, using its expression \eqref{pzz} in terms of sphere 4-point functions. Considering the corresponding set of sphere maps $\varphi(M)$ \eqref{phi}, we distinguish 3 cases:
\begin{itemize}
 \item If $mnp=0$, then the maps $M'\in\varphi(M)$ are characterized by their signatures. Each sphere 4-point function $Z_{M'}^*$ is the unique solution of crossing symmetry equations with spectrum $\mathcal{S}^*_{\sigma^{(x)}}$ in the $x$-channel, up to an overall constant. We can choose these constants such that the solutions $Z_{M'}^*$ are related by permutation symmetry \cite{nrj23}, and this determines $Z_M$ up to an overall constant via Eq. \eqref{pzz}. Requiring permutation symmetry ensures that contributions of odd spin fields cancel when we perform the sum over $M'$, therefore $Z_M$ has the expected spectrum $\mathcal{S}_{\sigma(M)}$.
 \item If $mnp\neq 0$ but $m,n,p$ are not all different, then the maps $M'$ come in pairs, for example $(m,m,p)^\pm$. Such maps are not characterized by their signatures, so we cannot compute $Z^*_{M'}$. However, we only need to compute combinations of the type $\sum_\pm Z^*_{(m,m,p)^\pm}$, where the two terms are related by a field permutation.
 Therefore, $\sum_\pm Z^*_{(m,m,p)^\pm}$ is the unique solution of crossing symmetry with
 \begin{align}
  \left(\mathcal{S}^{(s)},\mathcal{S}^{(t)},\mathcal{S}^{(u)}\right) =  \left(\left.\mathcal{S}^*_{2m}\right|_{rs\in 2\mathbb{Z}}, \mathcal{S}^*_{m+p},\mathcal{S}^*_{m+p}\right)\ ,
  \label{sss}
 \end{align}
such that 4-point structure constants obey the constraints from permutation invariance \cite{nrj23},
 \begin{align}
  D^{(s)} = (-)^{r s} D^{(s)} \quad , \quad D^{ (t)}_{(r,s)} = (-)^{rs} D^{(u)}_{(r,s)}\ .
 \end{align}
(The constraints on $D^{(s)}$ amount to restricting the spectrum to even spins.)
 This allows us to unambiguously determine $Z_M$.
 \item If $m,n,p$ are all different and nonzero, nothing in bootstrap equations can distinguish $Z_{(m,n,p)}$ from $Z_{(m,p,n)}$, and this ambiguity persists on the sphere. All we can do is to determine the 2-dimensional space of correlation functions that is spanned by these solutions. While this case is generic, it only occurs if $r_1\geq 6$, and we will not reach such high values of $r_1$ in the explicit examples of Section \ref{sec:data}.
\end{itemize}
In some cases, we can compute $Z_M$ directly on the torus, without using the relation with the sphere:
\begin{itemize}
\item Cases when $Z_M=0$ due to symmetries of the map $M$, see Eq. \eqref{som}.
 \item Cases when the torus map $M$ is characterized by its signature, see Eq. \eqref{mcbs}.
 \item Cases $M=(m,0,0)$, when the signature is zero but we can impose additional constraints on structure constants.
 In terms of the reduced structure constants of Section \ref{sec:sc}, these constraints are
 \begin{align}
  \frac12 \leq r \leq \frac{r_1-1}{2} \implies d_{(r,s)} = 0 \ .
 \end{align}
While inferred from results on the sphere \cite{nrj23}, these constraints can be imposed directly on the torus. The number of constraints is $\frac12r_1(r_1-1)\geq |\mathcal{M}(r_1)|-1$, which is enough for singling out a solution.
\end{itemize}
More generally, we can compute $Z_M$ on the torus as soon as we have enough information for singling out the combinatorial map $M$, i.e. as soon as we know $\mathcal{N}_{(r_1,s_1)}-1$ structure constants, where $\mathcal{N}_{(r_1,s_1)}$ \eqref{nroso} is the number of solutions of modular covariance for $\left<V_{(r_1,s_1)}\right>$.
These structure constants may come from numerical boostrap results on the sphere, or from other sources.

\section{Conformal bootstrap on the torus and on the sphere}

In this section we will describe the conformal bootstrap equations of critical loop models: modular covariance of torus 1-point functions, and crossing symmetry of sphere 4-point functions. These are linear equations whose coefficients are conformal blocks, and whose unknowns are structure constants: we will study such objects, and how they behave under the sphere-torus relation.

In principle, the known sphere-torus relation for Virasoro blocks implies the relation for logarithmic and interchiral blocks, for structure constants, and for correlation functions --- since all these objects are ultimately determined by Virasoro blocks. It is however useful and nontrivial to write these relations explicitly. This will allow us to check torus results by comparing them with the corresponding sphere results, and to compute torus 1-point functions from the sphere, whenever they cannot be computed directly on the torus.

\subsection{Structure constants}\label{sec:sc}

In critical loop models, we know all 3-point structure constants, but this is not enough for computing all correlation functions \cite{jnrr25}. While they do not give the full answer, sphere 3-point structure constants play a crucial role for determining sphere 4-point structure constants analytically \cite{nrj23}.

Sphere 3-point structure constants can be written as $C=\omega \check{C}$, with
\begin{align}
 \check{C}_{1,2,3} = \prod_{\epsilon_1,\epsilon_2,\epsilon_3=\pm} \Gamma_\beta^{-1} \left(q + \tfrac{\beta}{2}\left|\textstyle{\sum_i} \epsilon_ir_i\right| + \tfrac{\beta^{-1}}{2}\textstyle{\sum_i} \epsilon_is_i\right) \ , \quad \text{where}\quad q=\frac{\beta+\beta^{-1}}{2}\ ,
\end{align}
and $\Gamma_\beta$ is the Barnes double Gamma function. The sign factor $\omega \in
\{\pm1\}$ is characterized by its invariance under cyclic permutations, as well as the shift
equations \cite[(4.13a)]{rib24}
\begin{align}
 \frac{\omega_{1,2,3}\big|{}_{s_1\to s_1+2}}{\omega_{1,2,3}} &\ =\  (-)^{2r_3}(-)^{\max(2r_1, 2r_2, 2r_3,r_1+r_2+r_3)} \ ,
 \label{tpt}
 \\
 \frac{\omega_{1,2,3}\Big|{}_{\substack{s_1\to s_1+1 \\ s_3\to s_3+1}}}{\omega_{1,2,3}} &\underset{r_i\in\mathbb{N}^*}{=} (-)^{\max(r_3,r_2-r_1)} \ .
\label{tppt}
\end{align}

\subsubsection{Reference structure constants for torus 1-point and sphere 4-point functions}

From the 3-point structure constants, we build reference structure constants for the torus
1-point function $\left<V_1\right>$ and the sphere 4-point function
$\left<V_1V_2V_3V_4\right>^\text{sphere}$. We denote them as $\hat D_k$, using
\begin{itemize}
 \item the index $k$ for a generic field $V_{(r,s)}$,
 \item the index $I$ for the identity field $V^d_{\langle 1,1\rangle}$,
 \item the index $0$ for the diagonal field $V_{P_{(0,\frac12)}}$.
\end{itemize}
In the case of $s$-channel structure constants on the sphere, we write $\hat D^{(s)}_k$ for a generic 4-point function, and $\hat D^\text{sphere}_k$ for a 4-point function that corresponds to a torus 1-point function:
\begin{align}
\renewcommand{\arraycolsep}{18pt}
\renewcommand{\arraystretch}{1.8}
\begin{array}{ccc}
 \begin{tikzpicture}[baseline = (base), scale = .45]
   \coordinate (base) at (0, 0);
   \node[below] at (-2.3, 0) {$1$};
   \node[right] at (3, 0) {$k$};
   \draw[ultra thick] (-2.5, 0) to (0, 0);
  \draw[ultra thick] plot [smooth, tension = .8] coordinates {(0, 0) (2, 1) (3, 0) (2, -1) (0, 0)};
 \end{tikzpicture}
&
 \begin{tikzpicture}[baseline = (base), scale = .3]
   \coordinate (base) at (0, 0);
  \draw[ultra thick] (-1 ,2) -- (0, 0) -- (4, 0) -- (5, 2);
  \draw[ultra thick] (-1, -2) -- (0, 0);
  \draw[ultra thick] (5, -2) -- (4, 0);
  \node[left] at (-1, 2) {$2$};
  \node[right] at (5,2) {$3$};
  \node[right] at (5,-2) {$4$};
  \node[left] at (-.8,-2) {$1$};
  \node[below] at (2, 0) {$k$};
 \end{tikzpicture}
 &
 \begin{tikzpicture}[baseline = (base), scale = .3]
   \coordinate (base) at (0, 0);
  \draw[ultra thick] (-1 ,2) -- (0, 0) -- (4, 0) -- (5, 2);
  \draw[ultra thick] (-1, -2) -- (0, 0);
  \draw[ultra thick] (5, -2) -- (4, 0);
  \node[left] at (-1, -2) {$0$};
  \node[right] at (5,2) {$0$};
  \node[right] at (5,-2) {$0$};
  \node[left] at (-.8,2) {$1$};
  \node[below] at (2, 0) {$k$};
 \end{tikzpicture}
 \\
 \hat D_k = \frac{C_{1,k,k}}{C_{I,k,k}}  & \hat D^{(s)}_k = \frac{\check C_{1,2,k}\check C_{3,4,k}}{C_{I,k,k}} & \hat D^\text{sphere}_k = \frac{C_{0,1,k}C_{0,0,k}}{C_{I,k,k}}
 \end{array}
 \label{hd}
\end{align}
In the case of $C_{1,2,k}$ and $C_{3,4,k}$, there is no simple way to write sign factors that satisfy the shift equations \eqref{tpt} and \eqref{tppt}, so we use $\check C_{1,2,k}$ and $\check C_{3,4,k}$ instead.

Let us explicitly compute the reference structure constant $\hat D_k$, starting with sign factors. We focus on the dependence of the channel field $V_{(r,s)}$ on the second Kac index, and do not worry about shifts of $s_1$. We have $\omega_{1,k,k}\big|{}_{s\to s+2}=\omega_{1,k,k}$ and $\frac{\omega_{1,k,k}\big|{}_{s\to s+1}}{\omega_{1,k,k}} \underset{r\in\mathbb{N}}{=} (-)^{\max(r,r_1-r)}$. The solution of these shift equations may be written as $\frac{\omega_{1,k,k}}{\omega_{I,k,k}} = (-)^{r_1\lfloor s\rfloor\delta_{2r<r_1}}$. This leads to
\begin{align}
 \hat D_{(r,s)} = \frac{(-)^{r_1\lfloor s\rfloor\delta_{2r<r_1}}\prod_\pm \Gamma_\beta^2(\beta^{\pm 1}) \prod_{\pm,\pm} \Gamma_\beta(\beta^{\pm 1} + 2P_{(r,\pm s)})}{\prod_\pm \Gamma_\beta^2(q+P_{(r_1,\pm s_1)})
 \prod_{\pm}\prod_{\epsilon = \pm} \Gamma_\beta(q+ P_{(|r_1+2\epsilon r|,\pm (s_1+2\epsilon s))})} \ .
 \label{hdrs}
\end{align}
This formula also holds in the case $r=0$, except for the sign factor, which would not make sense for $s\in\mathbb{C}$:
\begin{align}
 \hat D_P = \frac{\prod_\pm \Gamma_\beta^2(\beta^{\pm 1})\prod_{\pm,\pm} \Gamma_\beta(\beta^{\pm 1} \pm 2P)}{\prod_\pm \Gamma_\beta^2(q+P_{(r_1,\pm s_1)})
 \prod_{\pm,\pm} \Gamma_\beta(q+ P_{(r_1,\pm s_1)}\pm 2P)} \ .
 \label{hdp}
\end{align}
Therefore, $\hat D_P$ only obeys the shift equation for $P\to P+\frac12\beta^{-1}$ up to a sign $(-)^{r_1}$. This does not affect the values of $\hat D_P$ on the diagonal spectrum \eqref{conv}, which is generated by shifts $P\to P+\beta^{-1}$.

To compute interchiral blocks, it is useful to know how these structure constants behave under shifts:
\begin{multline}
 \label{hds2}
 \frac{\hat D_{(r,s+1)}}{\hat D_{(r,s-1)}} = \prod_{a\in\{0,1,\beta^{-2},1-\beta^{-2}\}} \frac{\Gamma(a+r-\beta^{-2}s)}{\Gamma(a+r+\beta^{-2}s)} \times
 \\
 \prod_{a\in\left\{\frac{1-\beta^{-2}}{2},\frac{1+\beta^{-2}}{2}\right\}}\prod_\pm \frac{\Gamma\left(a+|r\pm \frac{r_1}{2}| +\beta^{-2}(s\pm  \frac{s_1}{2})\right)}{\Gamma\left(a+|r\pm \frac{r_1}{2}| -\beta^{-2}(s\pm  \frac{s_1}{2})\right)}\ ,
\end{multline}
\begin{multline}
  \label{hds1}
  \frac{\hat D_{(r,s+1)}}{\hat D_{(r,s)}}
  \underset{r\in\mathbb{N}}{=} (-)^{r_1\delta_{2r<r_1}}
  \frac{\prod_{a\in\{0,1-\beta^{-2}\}}\Gamma(a+r-\beta^{-2}s)}{\prod_{a\in\{1,\beta^{-2}\}}\Gamma(a+r+\beta^{-2}s)}
  \\
  \times
   \prod_{\pm}
    \frac{\Gamma \left( \frac{1+\beta^{-2}}{2} +|r\pm \frac{r_1}{2}| +\beta^{-2}(s\pm  \frac{s_1}{2}) \right)}{\Gamma \left( \frac{1-\beta^{-2}}{2} +|r\pm \frac{r_1}{2}| -\beta^{-2}(s\pm  \frac{s_1}{2}) \right)}\ .
\end{multline}

\subsubsection{Sphere-torus relation for reference structure constants}

Let us compare $\hat D_{(r,s)}$ with $\hat D_{(2r,s)}^\text{sphere}$. We first compare sign factors. We determine the sign factors of $\hat D_{(2r,s)}^\text{sphere}$ \eqref{hd}, by solving the equations \eqref{tpt} and \eqref{tppt}.
Notice that the field $V_0=V_{(0,\frac12)}=V_{(0,-\frac12)}$ is invariant under a shift of its second Kac index by one unit.
The sign factors $\omega_{0,0,(2r,s)}=\omega_{I,(2r,s),(2r,s)}=1$ are trivial, since the channel field $V_{(2r,s)}$ has a first Kac index $2r\in 2\mathbb{N}$. It remains to check $\omega_{0,1,(2r,s)} = \frac{\omega_{1,(r,s),(r,s)}}{\omega_{I,(r,s),(r,s)}}$, which holds. (Since our sign factors depend on $s$ but not on $s_1$, we did not need to apply the sphere-torus relation to the field $V_{(r_1,s_1)}$.)

Then let us compare factors that involve double Gamma functions. We need to compare values of $\Gamma_\beta$ and $\Gamma_{\beta'}$, where $\beta'=\frac{\beta}{\sqrt{2}} $. Using Eq. \cite[eq. (B.28)]{eber23}, we find
\begin{align}
 \Gamma_\beta\left(q+z\right) = 2^{-\frac14 z^2} \prod_\pm \frac{\Gamma_{\beta'}(q'\pm \frac{\beta'^{-1}}{4}+\frac{z}{\sqrt{2}})}{\Gamma_{\beta'}(q'\pm \frac{\beta'^{-1}}{4})}\ .
\end{align}
This allows us to compute the ratio of $\hat D_{(r,s)}$ with $\hat D_{(2r,s)}^{\text{sphere}}{}'$, where the prime indicates that the values of the implicit variables $\beta$ and $(r_1,s_1)$ are given by the sphere-torus relation of Table \eqref{rel}:
\begin{align}
\hspace{-5mm}
 \frac{\hat D_{(r,s)}}{\hat D_{(2r,s)}^{\text{sphere}}{}'} = 16^{2\Delta+2\bar\Delta}
 \frac{\sqrt{2}^{\Delta_1+\bar\Delta_1}}{\sqrt{2}^{(\beta-\beta^{-1})^2}} \prod_\pm \Gamma_{\beta}^{-2} \big( q + \tfrac{\beta}{2} r_1 \pm \tfrac{\beta^{-1}}{2}s_1 \big) \frac{\Gamma_{\beta'}(q'\pm \frac{\beta'}{2})}{\Gamma_{\beta'}(q'\pm \frac{\beta'^{-1}}{4})}
 \, .
 \label{dpd}
\end{align}
Only the first factor $16^{2\Delta+2\bar\Delta}$ depends on the channel field $V_{(r,s)}$: this factor will cancel a prefactor in the sphere-torus relation for conformal blocks \eqref{faf}. The rest of the ratio depends only on the central charge and the external field $V_1$, and plays the role of an unimportant normalization in $\left<V_1\right>$.

\subsubsection{Residues of structure constants}

Combinatorial maps with signatures of the type $(m,0,0)$ allow topologically nontrivial closed loops, and the corresponding correlation function depends on the weight $w(P)$ \eqref{wP} of such loops. See the second column of Table \eqref{stmaps} for an example with $m=3$. The structure constants $D_k$ then depend on $P$. We will now discuss their poles and residues as functions of $P$, along the lines of \cite{rib24}. Much of the discussion will be valid both on the torus and on the sphere.

Consider the structure constant $D_P$ for a diagonal channel field $V_P$.
Motivated by analogous results for diagonal fields on the sphere \cite{nrj23} and on the disc \cite{djnrs25}, which generalize the Delfino--Viti ansatz for 3-point connectivities \cite{dv10},
we conjecture that $D_P$ coincides with its reference value:
\begin{align}
 D_P = \hat D_P\ .
\end{align}
The corresponding term in the expansion \eqref{vdg} of the correlation function is $\hat D_P \mathcal{G}_P=\hat D_P\left|\mathcal{F}_P\right|^2=\hat D_P \mathcal{F}_P\overline{\mathcal{F}}_P$, where $\mathcal{F}_P$ and $\overline{\mathcal{F}}_P$ are left-moving and right-moving Virasoro blocks respectively.
This term
has poles for $P=P_{(r,s)}$ with $(r,s)\in\mathbb{N}^*\times \mathbb{Z}$. But we expect that the correlation function
itself is the critical limit of a polynomial function of $w(P)$, as computed by a lattice loop
model. Therefore, we conjecture that the correlation function is holomorphic in $w(P)$, and has
no poles.  To cancel the poles of $\hat D_P\left|\mathcal{F}_P\right|^2$, the structure constants
$D_{(r,s)}$ of the non-diagonal sector must themselves have poles, with the residues
\begin{align}
 \underset{w=w(P_{(r,s)})}{\operatorname{Res}} D_{(r,s)}(w)\ \underset{r,s\in\mathbb{N}^*}{=}\ - \left[(-2\pi\beta)\frac{\sin(2\pi\beta P_{(r,s)})}{P_{(r,s)}}\right] \overline{R}_{r,s}\hat D_{P_{(r,s)}}\ .
 \label{rdrd}
\end{align}
On the right-hand side, the factor between brackets is the Jacobian of the change of variables from the conformal dimension $\Delta$ to the loop weight $w$, and $R_{r,s}$ is
the right-moving version of the Virasoro block residue defined by
\begin{align}
\forall r,s\in\mathbb{N}^*\ , \qquad
\underset{\Delta=\Delta_{(r,s)}}{\operatorname{Res}} \mathcal{F}_\Delta = R_{r,s}\mathcal{F}_{\Delta_{(r,-s)}}\ ,
\label{rfd}
\end{align}
and $\overline{R}_{r,s}$ is the same quantity evaluated for right-moving momentums $\bar{P}_i$.
The explicit expression of $R_{r,s}$ depends on whether we are dealing with a torus 1-point block or a sphere $s$-channel 4-point block:
\begin{align}
 R^\text{torus}_{r,s} = \frac{c^{2r,2s}(P_1)}{b^{r,s}} \quad , \quad R^{(s)}_{r, s} = \frac{\prod_\pm c^{r,s}(P_1\pm P_2)c^{r,s}(P_3\pm P_4)}{b^{r,s}} \ ,
 \label{rrs}
\end{align}
where we define
\begin{align}
 c^{r,s}(P) &= \frac{\prod_{\pm}\Gamma_\beta\left(q +P \pm P_{(r,s)}\right)}
 {\prod_{\pm}\Gamma_\beta\left(q +P \pm P_{(r,-s)}\right)} \ ,
 \\
 b^{r,s} &=
 \frac{-\prod_\pm \Gamma_\beta\left(\beta \pm 2P_{(r,s)}\right)}{P_{(r,s)}\Gamma_\beta\left(\beta+2P_{(r,-s)}\right)\operatorname{Res}_{\beta-2P_{(r,-s)}}\Gamma_\beta}\ .
\end{align}
The residues $R^\text{sphere}_{r,s}$ that appear in $\left<V_0V_1V_0V_0\right>^\text{sphere}$ obey
\begin{align}
R^\text{sphere}_{2r,s}{}' = 2\cdot 16^{-2rs}R_{r,s}^\text{torus}
\quad , \quad
R^\text{sphere}_{2r-1,s} =\ 0 \ .
 \label{rtrs}
\end{align}
Together with the identity $\Delta_{(r,s)}+rs=\Delta_{(r,-s)}$, this allows us to check that the residues \eqref{rdrd} obey the same sphere-torus relation \eqref{dpd} as the structure constants themselves.

Let us now compute the ratio of the residue by the reference structure constant:
\begin{align}
  \label{eq:rhors}
 \Xi_{r,s} \equiv \frac{\underset{w=w(P_{(r,s)})}{\operatorname{Res}} D_{(r,s)}(w)}{\hat D_{(r,s)}}\ \  \implies\ \  \Xi_{r,s}^\text{torus}=
 \frac{\Theta_1^{r,s}}{\kappa_{r,s}} \ \ , \ \ \Xi_{r,s}^{(s)} = \omega^{(s)}_{r,s}\frac{\rho^{r,s}_{1,2}\rho^{r,s}_{3,4}}{\kappa_{r,s}}\ ,
\end{align}
where $\omega^{(s)}_{r,s}\in\{-1,1\}$ is a sign factor, and we define
\begin{align}
  \Theta_1^{r,s} &=  - (-)^s(-)^{r_1s\delta_{2r>r_1}}  \frac{\check C_{(r_1,s_1),P_{(r,s)},P_{(r,s)}}}{ \check C_{(r_1,s_1),(r,s),(r,s)}}c^{2r,2s}(\bar P_1)\ ,
  \\
  \rho^{r,s}_{1,2} &=  \frac{\check C_{P_{(r,s)},(r_1,s_1),(r_2,s_2)}}{\check C_{(r,s),(r_1,s_1),(r_2,s_2)}}\prod_\pm c^{r,s}(\bar P_1\pm \bar P_2) \ ,
  \\
  \kappa_{r,s} &= \frac{b^{r,s}P_{(r,s)}}{(-2\pi\beta)\sin(\pi\beta^2r)} \frac{ \check C_{I,P_{(r,s)},P_{(r,s)}}}{\check C_{I,(r,s),(r,s)}}\ ,
  \label{dkrs}
\end{align}
where we used $\sin(2\pi\beta P_{(r,s)}) = (-)^s\sin(\pi\beta^2r)$.
Writing $P=P_{(r,s)}$, the factors that appear in $\Theta_1^{r,s}$ read
\begin{align}
 c^{2r,2s}(\bar P_1)
 &= \frac{\prod_\pm \Gamma_\beta(q +\bar P_1 \pm 2P)}{\prod_\pm \Gamma_\beta(q +\bar P_1 \pm 2\bar P)} \ ,
 \\
 \check C_{(r_1,s_1),P,P} & =\prod_\pm \Gamma_\beta^{-1}\left(q+P_1\pm 2P\right)\Gamma_\beta^{-1}\left(q+\bar P_1\pm 2P\right)\ ,
 \\
 \check C_{(r_1,s_1),(r,s),(r,s)} &=
 \renewcommand{\arraystretch}{1.4}
 \left\{\begin{array}{ll}
       \prod_\pm  \Gamma_\beta^{-1}\left(q\pm P_1 + 2P\right)\Gamma_\beta^{-1}\left(q\pm \bar P_1 + 2\bar P\right) & \text{ if }2r\geq r_1 \ ,
       \\
       \prod_\pm  \Gamma_\beta^{-1}\left(q+ P_1 \pm 2P\right)\Gamma_\beta^{-1}\left(q+ \bar P_1 \pm 2\bar P\right) & \text{ if }2r\leq r_1 \ .
        \end{array}
\right.
\end{align}
We deduce
\begin{align}
 \Theta^{r,s}_1 = (-)^{r_1s+s+1} \prod_{j\overset{1}{=}\frac{r_1+1}{2}-r}^{r-\frac{r_1+1}{2}} 2\cos\pi\left(j\beta^2 -\tfrac{s_1}{2}\right)\ .
 \label{dors}
\end{align}
This is a product of $2r-r_1$ factors, where the index $j$ runs by increments of $1$. For $2r\leq r_1$ we have $\Theta^{r,s}_1=(-)^{r_1s+s+1}$. On the sphere, the comparable quantity is \cite[Eq. (2.52)]{nrj23}
\begin{align}
 \rho^{r,s}_{1,2}= (-)^{s\min(r,|r_1-r_2|)\delta_{r_1<r_2}} \prod_\pm \prod_{j\overset{1}{=}-\frac{r-1-|r_1\pm r_2|}{2}}^{\frac{r-1-|r_1\pm r_2|}{2}} 2\cos\pi\left(j\beta^2+\tfrac{s-s_1\mp s_2}{2}\right)\ .
\label{rhors}
\end{align}

\subsubsection{Polynomial denominators of residues}

The denominator $\kappa_{r,s}$ \eqref{dkrs} of the residue is a priori defined only for $r,s$ integer: for $s$ fractional or $r$ half-integer, the structure constant $D_{(r,s)}(P)$ does not have poles. Nevertheless, it is useful to define $\kappa_{r,s}$ for all values of $r,s$, because the ratio $\frac{D_{(r,s)}}{\hat D_{(r,s)}}$ turns out to be a rational function of loop weights, whose denominator is $\kappa_{r,s}$ (except for the pole term if present), see Eq. \eqref{eq:15}. The general expression for $\kappa_{r,s}$, which reduces to the expression \eqref{dkrs} for $r,s\in\mathbb{N}^*$, is
\begin{align}
 \kappa_{r,s} \ \underset{r\geq 1}{=}\ \frac{2^{2\lfloor r + \frac12 \rfloor -1}}{\sin\pi\left(r-\lfloor r\rfloor+s\right)} \prod_{j\overset{1}{=} 1-r}^{r-1} \sin \pi(\beta^2 j +s) \ .
 \label{kappa}
\end{align}
This is a polynomial function of the loop weight $n$ \eqref{nb}, with
\begin{align}
\renewcommand{\arraystretch}{1.2}
 \deg_n\kappa_{r,s} = \left\{\begin{array}{ll} r(r-1) & \text{ if } r\in \mathbb{N}^*\ ,
                                \\ (r-\frac12)^2 & \text{ if } r\in \mathbb{N}+\frac32\ .
                               \end{array}\right.
\label{degk}
\end{align}
Let us write manifestly polynomial expressions for $\kappa_{r,s}$:
\begin{align}
 \kappa_{r,s} \ &\underset{r\in \mathbb{N}^*}{=}\  2\prod_{j=1}^{r-1}\prod_{k=0}^{j-1} \left(n^2 -4\cos^2\pi \tfrac{k+s}{j}\right) \ ,
 \\
 \kappa_{r,s} \ &\underset{r\in \mathbb{N}+\frac32}{=}\ \frac{2(-)^{r-\frac12}}{\cos\pi s} \prod_{j\overset{2}=1}^{2r-2} \prod_{k=0}^{j-1}\left(n+2\cos 2\pi \tfrac{k+s}{j}\right)\ ,
\end{align}
where the product over $j$ runs over odd integers in the second formula.
Let us write the first few examples, while completing the definition \eqref{kappa}, which did not cover the cases $r=0,\frac12$. We use the golden ratio $\varphi = \frac{1+\sqrt{5}}{2}= 2\cos(\frac{\pi}{5})$:
\begin{subequations}
 \begin{align}
 \kappa_{(0,s)} &=\frac{1}{2\sin^2(\pi s)} \quad , \quad \kappa_{(\frac12, s)} = 2 \quad ,\quad \kappa_{(1,s)} = 2\ ,
 \\
 \kappa_{(\frac32,0)} &= -2(n+2) \quad , \quad \kappa_{(\frac32,\frac23)} = 4(n-1)\ ,
 \\
 \kappa_{(2,0)} &= 2(n^2-4) \quad , \quad \kappa_{(2,\frac12)} = 2n^2\ ,
 \\
 \kappa_{(\frac52, 0)} &= 2(n-1)^2(n+2)^2\ ,
\\
 \kappa_{(\frac52,\frac25)} &= 4(n-\varphi)(n+\varphi^{-1})(\varphi n^2-n-1-2\varphi)\ ,
 \\
 \kappa_{(\frac52,\frac45)} &= -4(n-\varphi)(n+\varphi^{-1})(\varphi^{-1}n^2+n+1-2\varphi^{-1})\ ,
 \\
 \kappa_{(3,0)} &= 2n^2(n^2-4)^2 \quad , \quad \kappa_{(3,\frac13)} = 2(n^2-1)^2(n^2-3)\ .
\end{align}
\end{subequations}

\subsubsection{Sphere-torus relation for polynomial factors of structure constants}

Since residues are deduced from reference structure constants, the sphere-torus relation for the latter implies a sphere-torus relation for the former. And indeed, we find the following relations for the factors that appear in the residues \eqref{rhors}:
\begin{align}
 \frac{\kappa'_{2r,s}}{\kappa_{r,s}} =  \rho'^{\, 2r,s}_{0,0}\qquad , \qquad
 \Theta_1^{r,s} = -(-)^{r_1s\delta_{r_1>2r}}\rho'^{\, 2r,s}_{0,1}\ .
\end{align}
These relations make sense for $(r,s)\in \mathbb{N}^*\times \mathbb{Z}$, but we can generalize the first relation to generic $r,s$, and we find
\begin{align}
 \frac{\kappa'_{2r,s}}{\kappa_{r,s}} = \frac{\sin\pi\left(r-\lfloor r\rfloor+s\right)}{\sin \pi s}4^{\lceil r-\frac12\rceil} \prod_{j\overset{1}{=}\frac12-r}^{r-\frac12} \sin\pi(\beta^2j+s)\ .
\end{align}
For a sphere 4-point function, the counterpart of our expression \eqref{eq:15} for a structure constant in terms of a polynomial $d_{(r,s)}$ is \cite[Eq. (5.33)]{rib24}:
\begin{align}
 D_{(r,s)}^M =  \frac{\hat D_{(r,s)}^{(s)}}{\kappa^{r,s}} \left(d_{(r,s)}^M  - \delta_{\sigma,0}\delta_{s\in\mathbb{Z}}\frac{(-)^{(r+1)s}\rho^{r,s}_{1,2}\rho^{r,s}_{4,3}}{w-w_{(r,s)}}\right) \ .
 \label{sdrs}
\end{align}
Here we explicitly indicated the combinatorial map $M$, with $\sigma$ its signature. The relation \eqref{pzz} for solutions of bootstrap equations leads to
\begin{align}
 d_{(r,s)}^M = (-)^{r_1\lfloor s\rfloor\delta_{2r<r_1}} r\frac{\kappa_{r,s}}{\kappa'_{2r,s}}\cdot \frac{1}{|\varphi(M)|}\sum_{M'\in\varphi(M)} d'^{\, (x),M'}_{(2r,s)}\ ,
\end{align}
where $x\in\{s,t,u\}$ is some channel, and the set of maps $\varphi(M)$ is given in Eq. \eqref{phi}. (Depending on $M$, part or all of the sum over $M'$ might be traded for a sum over $x$.)
In this relation, it is not manifest that $d_{(r,s)}$ is polynomial if $d'^{\, (x)}_{(2r, s)}$ is. The conjecture that $d_{(r,s)}$ is polynomial implies that the sum over channels has a factor $\frac{\kappa'_{2r,s}}{\kappa_{r,s}}$, which is a polynomial function of $n$ of degree $\lfloor 2r^2-\frac12\rfloor$. We find that the sphere results \cite{nrj23} fulfil this prediction.

\subsection{Conformal blocks}

\subsubsection{4-point Virasoro blocks on the sphere}

Let us review a few features of the Virasoro blocks that appear in sphere 4-point functions $\left<V_1V_2V_3V_4\right>^\text{sphere}$. (More details are found in \cite{rib24}.)
The $s$-channel blocks take the form
\begin{align}
  \label{Fps}
 \mathcal{F}^{(s),\text{sphere}}_P(q) = (16q)^{P^2} A(q) H_P^{\text{sphere}}(q)\ ,
\end{align}
where $H^{\text{sphere}}_P(q)$ is a power series such that $H^{\text{sphere}}_P(0)=1$. From Zamolodchikov's recursion \eqref{rfd}, this power series obeys
\begin{align}
 \underset{\Delta=\Delta_{(r,s)}}{\operatorname{Res}} H^{\text{sphere}}_\Delta(q) &= R^{\text{sphere}}_{r,s}(16q)^{rs} H^{\text{sphere}}_{\Delta_{(r,-s)}}(q)\ .
\end{align}
The prefactor $A(q)$ is expressed in terms of Jacobi theta functions $\theta_k(q)$ as
\begin{align}
 A(q) = \prod_{k=2,3,4} \theta_k^{-4}(q)^{P_k^2+(-)^k\Delta_1}\ ,
 \label{Aq}
\end{align}
where $P_k$ is the momentum of the field $V_k$.
The nome $q$ is related to the cross-ratio $z$ of the sphere, and it is convenient to introduce a modulus $\tau$ such that
\begin{align}
z = \frac{\theta_2^4(q)}{\theta_3^4(q)} \quad , \quad 1-z =\frac{\theta_4^4(q)}{\theta_3^4(q)}
\quad , \quad
 q \underset{\text{sphere}}{=} e^{i\pi\tau} \ .
 \label{zqt}
\end{align}
Crossing symmetry equations also involve $t$-channel and $u$-channel blocks, which may be obtained from $s$-channel blocks by certain permutations of the 4 fields. Let us indicate how the relevant permutations act on the geometric variables $z,\tau,\theta_k(q)$, on the momentums $P_i$, and on the prefactor $A(q)$:
\begin{align}
 \renewcommand{\arraystretch}{1.4}
 \begin{array}{|c|c|c|r|c|c|}
  \hline
  s  & z & \tau & (\theta_2^4,\theta_3^4,\theta_4^4) & (P_1, P_2, P_3,P_4)  & A
  \\
  \hline
  t & 1-z & -\frac{1}{\tau} & \tau^2(\theta_4^4,\theta_3^4,\theta_2^4) & (P_1,P_4,P_3,P_2) & \tau^{-2\delta} A
  \\
  \hline
  u & \frac{1}{z} & \frac{\tau-2}{\tau-1} & (\tau-1)^2(-\theta_3^4,-\theta_2^4,\theta_4^4) & (P_1,P_3,P_2,P_4) & (\tau-1)^{-2\delta} (-)^{P_2^2+P_3^2} A
  \\
  \hline
 \end{array}
 \label{tab:stu}
\end{align}
where we defined the exponent
\begin{align}
 \delta = \sum_{k=1}^4 P_k^2 - P_{(1,1)}^2\ .
\end{align}
This means that the $t$- and $u$-channel blocks are
\begin{align}
 \mathcal{F}^{(t),\text{sphere}}_P(q) &=  \tau^{-2\delta} A(q) \left[(16q)^{P^2}H^{\text{sphere}}_P(q)\right]_{\substack{P_2\leftrightarrow P_4 \\ z\leftrightarrow 1-z}} \ ,
 \\
 \mathcal{F}^{(u),\text{sphere}}_P(q) &= (\tau-1)^{-2\delta} (-)^{P_2^2+P_3^2} A(q) \left[(16q)^{P^2}H^{\text{sphere}}_P(q)\right]_{\substack{P_2\leftrightarrow P_3 \\ z\leftrightarrow \frac{1}{z}}} \ .
\end{align}

\subsubsection{1-point Virasoro blocks on the torus}

The Virasoro blocks that appear in a torus 1-point function $\left<V_1\right>$ take the form \cite{fl09}
\begin{align}
 \mathcal{F}_P(q) = \frac{q^{P^2}}{\eta(q)}H_P(q)\ .
\end{align}
Here $\eta(q)=q^{\frac{1}{24}}\prod_{n=1}^\infty (1-q^n)$ is the Dedekind eta function, and $H_P(q)$ is a power series in $q$ such that $H_P(0)=1$, again determined by Zamolodchikov's recursion \eqref{rfd}:
\begin{align}
 \underset{\Delta=\Delta_{(r,s)}}{\operatorname{Res}} H_\Delta(q) &= R_{r,s}q^{rs} H_{\Delta_{(r,-s)}}(q)  \ .
\end{align}
The relation between the parameter $q$ and the modulus $\tau$ of the torus slightly differs from the relation \eqref{zqt} between the sphere parameters of the same names,
\begin{align}
 q \underset{\text{torus}}{=} e^{2\pi i\tau} \ .
 \label{qt}
\end{align}
For any $g=\left(\begin{smallmatrix} a & b \\ c & d \end{smallmatrix}\right)\in PSL_2(\mathbb{Z})$, we define the $g$-channel Virasoro block
\begin{align}
 \mathcal{F}^{(g)}_P(q) = (c\tau +d)^{-\Delta_1} \mathcal{F}_P\left(e^{2\pi i \frac{a\tau+b}{c\tau+d}}\right)\ .
\end{align}
Modular covariance of a torus 1-point function is the equality of its decompositions into $g$-channel blocks for any $g\in PSL_2(\mathbb{Z})$. In particular, we rename $s,t,u$ the 3 channels associated to $g=
\left(\begin{smallmatrix} 1 & 0 \\ 0 & 1\end{smallmatrix}\right)$,
$\left(\begin{smallmatrix} 0 & 1 \\ -1 & 0\end{smallmatrix}\right)$,
and $
\left(\begin{smallmatrix} 1 & -2 \\ 1 & -1\end{smallmatrix}\right)$ respectively, so that
\begin{align}
 \mathcal{F}^{(s)}_P(q) &= \mathcal{F}_P(q)\quad , \quad \mathcal{F}^{(t)}_P(q) = \tau^{-\Delta_1}\mathcal{F}_P\left(e^{2\pi i (-\frac{1}{\tau})}\right) \ ,
 \nonumber
 \\
 \mathcal{F}^{(u)}_P(q) &= (\tau-1)^{-\Delta_1} \mathcal{F}_P\left(e^{2\pi i(\frac{\tau-2}{\tau-1})}\right)\ .
 \label{fstu}
\end{align}

\subsubsection{Sphere-torus relation for Virasoro blocks}

In the case of 4-point functions of the type $\left<V_0V_1V_0V_0\right>^\text{sphere}$, the residues $R^\text{sphere}_{r,s}$ are related to the torus residues $R_{r,s}$ by Eq. \eqref{rtrs}, and this implies
\begin{align}
 H_\Delta(q^2) = H^\text{sphere}_{\Delta'}{}'(q) \ .
\end{align}
In particular, the factor $2$ in Eq. \eqref{rtrs} cancels with the Jacobian of the map $\Delta\mapsto \Delta'$. This leads to the sphere-torus relation for blocks in any channel $x\in\{s,t,u\}$,
\begin{align}
 \mathcal{F}_P^{(x)}(\beta,P_1|q^2)= 16^{-2P^2} a(\beta,P_1|q) \mathcal{F}^{(x),\text{sphere}}_{\sqrt{2}P}\big(\tfrac{\beta}{\sqrt{2}},\tfrac{P_1}{\sqrt{2}}\big|q\big) \ .
 \label{faf}
\end{align}
So the sphere-torus relation keeps the modulus $\tau$ unchanged, while changing the nome $q$. And indeed the relation between $q$ and $\tau$ is not the same on the sphere \eqref{zqt} as on the torus \eqref{qt}. The relation involves the channel-independent prefactor
\begin{align}
 a(\beta,P_1|q) = \frac{1}{\eta(q^2) A^\text{sphere}{}'(q)} = z^{-\frac{1}{4\beta^2}} (1-z)^{\frac12-\frac{\beta^2}{8}-\frac{1}{4\beta^2}} \frac{\theta_2(q)^{2\Delta_1+1}}{\eta(q^2)}\ ,
 \label{aq}
\end{align}
where $A^\text{sphere}{}'(q)$ is the function $A(q)$ \eqref{Aq}, evaluated for $\left<V'_0V'_1V'_0V'_0\right>^\text{sphere}$, with variables dictated by the relation \eqref{rel}. Its channel-independence follows from the modular properties of the eta function,
\begin{align}
 \eta(-\tfrac{1}{\tau}) = \sqrt{-i\tau}\eta(\tau)\ \ , \ \  \eta(\tau+1)=e^{i\frac{\pi}{12}}\eta(\tau)\ \ , \ \  \eta(\tfrac{\tau-2}{\tau-1})=\sqrt{-i(\tau-1)}\eta(\tau)\ ,
 \label{eta}
\end{align}
and the modular properties of $A(q)$ \eqref{tab:stu}, where the exponent $\delta$ becomes $\delta'= \frac12\Delta_1+\frac14$.

\subsubsection{1-point logarithmic blocks on the torus}

If $r\notin \mathbb{N}^*$ or $s\notin \mathbb{Z}^*$, the primary field $V_{(r,s)}$ generates a Verma module of the conformal algebra. However,
if $r,s\in\mathbb{N}^*$, then both primary fields $V_{(r,s)}$ and $V_{(r,-s)}$ belong to the same logarithmic representation of the conformal algebra \cite{nr20}.
This leads to the conformal blocks
\begin{subequations}
\label{grs}
\begin{align}
\mathcal{G}_{(r,s)}&= \mathcal{F}_{\Delta_{(r,s)}}\mathcal{F}_{\Delta_{(r,-s)}} \ ,  & \text{(generic case)}
\\
 \mathcal{G}_{(r,s)} &\underset{r,s\in\mathbb{N}^*}{=} \frac{2P_{(r,s)}}{\overline{R}_{r,s} D_{P_{(r,s)}}} \left[ \operatorname{Res}_{P=P_{(r,s)}} + \operatorname{Res}_{P=P_{(r,-s)}}\right] D_P\left|\mathcal{F}_{P}\right|^2 \, ,  & \text{(logarithmic)}
 \\
 \mathcal{G}_{(r,s)} &\underset{r,-s\in\mathbb{N}^*}{=} 0 \ . & \text{(convention)}
\end{align}
\end{subequations}
To evaluate logarithmic blocks, we introduce the expansion of a Virasoro block near one of its poles,
\begin{align}
 \mathcal{F}_P = \frac{R_{r,s}}{2P_{(r,s)}(P-P_{(r,s)})} \mathcal{F}_{P_{(r,-s)}} + \mathcal{F}^\text{reg}_{P_{(r,s)}} + O(P-P_{(r,s)}) \ .
 \label{freg}
\end{align}
This leads to
\begin{multline}
  \mathcal{G}_{(r,s)} = \left(\mathcal{F}^\text{reg}_{P_{(r,s)}} -\tfrac{R_{r,s}}{2P_{(r,s)}} \mathcal{F}'_{P_{(r,-s)}}\right) \overline{\mathcal{F}}_{P_{(r,-s)}}
  + \frac{R_{r,s}}{\overline{R}_{r,s}} \mathcal{F}_{P_{(r,-s)}} \left(\overline{\mathcal{F}}^\text{reg}_{P_{(r,s)}} -\tfrac{\overline{R}_{r,s}}{2P_{(r,s)}} \overline{\mathcal{F}}'_{P_{(r,-s)}}\right)
  \\
  - \tfrac{R_{r,s}}{2P_{(r,s)}} \ell_{(r,s)}
  \left|\mathcal{F}_{P_{(r,-s)}}\right|^2
  \ ,
  \label{ggrs}
\end{multline}
where derivatives are with respect to the momentum $P$, and we introduced the coefficient
\begin{align}
 \ell_{(r,s)}= -\frac{D'_{P_{(r,s)}}}{D_{P_{(r,s)}}} + \lim_{P\to P_{(r,-s)}}\left[\frac{2}{P-P_{(r,-s)}} +\frac{D'_P}{D_P}\right]\ .
\end{align}
This formula is equally valid for sphere 4-point blocks and torus 1-point blocks. However, the particular expression of $\ell_{(r,s)}$ depends on the case. The expression for 4-point blocks is known \cite{nr20}. For 1-point blocks, we use the reference structure constant Eq. \eqref{hdp},
which leads to
\begin{multline}
  \beta \ell_{(r, s)} =2\sum_{\pm,\pm} \sum_{j=\frac12-s}^{s-\frac12} \psi\left(\tfrac{r_1+1}{2}\pm r\pm \tfrac{s_1}{2}\beta^{-2}  +j\beta^{-2}\right)
 \\
 -4\sum_\pm\sum_{j=1-s}^{s} \psi\left(\pm r+j\beta^{-2}\right)-4\pi\cot(\pi s\beta^{-2})\ ,
\end{multline}
where $\psi =(\log \Gamma)'$ is the digamma function, and for $r\in\mathbb{N}$ we define its value at its pole as $\psi(-r)=\psi(r+1)$.

\subsubsection{1-point interchiral blocks on the torus}

An interchiral block is an infinite linear combination of conformal blocks, related by shifts $s\to s+2$ of the channel field's second Kac index \cite{rib24}. In the case of 1-point blocks on the torus, the combination is
\begin{align}
 \widetilde{\mathcal{G}}_{(r,s_0)} = \sum_{s\in s_0+2\mathbb{Z}} \frac{\hat D_{(r,s)}}{\hat D_{(r,s_0)}} \mathcal{G}_{(r,s)}\ ,
\end{align}
where the relevant ratios of structure constants are given in Eq. \eqref{hds2}.

If $r\in\mathbb{N}$, the reference structure constants $\hat D_{(r,s)}$ \eqref{hd} also obey a shift equation \eqref{hds1} for $s\to s+1$. This is because $\hat D_{(r,s)}$ is written in terms of reference 3-point structure constants where the channel field appears twice.
We could therefore combine $\widetilde{\mathcal{G}}_{(r,s_0)}$ and $\widetilde{\mathcal{G}}_{(r,s_0+1)}$ into a single interchiral block. However, we numerically observe that the structure constant $D_{(r,s)}$ does not always obey the same shift equation as $\hat D_{(r,s)}$: in other words, we have in general $d_{(r,s+1)}\neq d_{(r,s)}$. We distinguish two cases:
\begin{itemize}
 \item For a map of the type $(r_1,0,0)$, unlike $\hat D_{(r,s)}$, $D_{(r,s)}$ depends on the weight $w$ of the nontrivial closed loops, and shifts flip the sign of $w$. Moreover, the diagonal reference structure constant $\hat D_P$ \eqref{hdp} only obeys the shift equation up to a sign $(-)^{r_1}$. This explains why we find in this case
 \begin{align}
  d_{(r,s+1)}(w) = (-)^{r_1} d_{(r,s)}(-w)\ .
  \label{dddd}
 \end{align}
And the pole term in \eqref{eq:15} behaves in the same way.
\item For other combinatorial maps, we find $d_{(r,s+1)} = \epsilon d_{(r,s)}$, with a sign $\epsilon\in\{-1,+1\}$ that depends only on the map (and not on $s_1,r,s$). Knowing $\epsilon$, we could build interchiral blocks $\widetilde{\mathcal{G}}_{(r,s_0)}+\epsilon\frac{\hat D_{(r,s_0+1)}}{\hat D_{(r,s_0)}}\widetilde{\mathcal{G}}_{(r,s_0+1)}$. We refrain from doing so for the sake of simplicity.
\end{itemize}

\subsubsection{Sphere-torus relation for conformal blocks}

Conformal blocks \eqref{grs} are assembled from structure constants and Virasoro blocks. They obey a non-chiral version of the sphere-torus relation for Virasoro blocks \eqref{faf},
\begin{align}
 \mathcal{G}_{(r,s)}^{(x)}\big(\beta,(r_1,s_1)\big|q^2\big)= \left| 16^{-2P^2} a(\beta,P_{1}|q)\right|^2 \mathcal{G}^{(x),\text{sphere}}_{(2r,s)}\big(\tfrac{\beta}{\sqrt{2}},(r_1,\tfrac{s_1}{2})\big|q\big) \ ,
 \label{gag}
\end{align}
where the modulus squared means multiplying left-moving and right-moving quantities, and $a(\beta,P_{1}|q)$ was given in Eq. \eqref{aq}.
The same relation holds for interchiral blocks.

\subsection{Modular covariance and crossing symmetry}\label{mccs}

For the sphere 4-point function $\left<V_1V_2V_3V_4\right>^\text{sphere}$, crossing symmetry equations read
\begin{align}
 \forall z \in\mathbb{C},\ \forall x_1\neq x_2 \in \{s, t, u\}, \quad \sum_{k\in\mathcal{S}^{(x_1)}} D^{(x_1)}_k \mathcal{G}^{(x_1)}_k(z)  = \sum_{k\in\mathcal{S}^{(x_2)}} D^{(x_2)}_k \mathcal{G}^{(x_2)}_k(z)
 \label{cross}
\end{align}
The unknowns are $\{D^{(x)}_k\}_{k,x}$.
For the torus 1-point function $\left<V_1\right>$, modular covariance equations read
\begin{align}
 \forall \tau \in \mathbb{C},\ \forall g_1\neq g_2 \in PSL_2(\mathbb{Z})  , \quad \sum_{k\in \mathcal{S}} D_k \mathcal{G}^{(g_1)}_k(\tau)  =  \sum_{k\in \mathcal{S}} D_k \mathcal{G}^{(g_2)}_k(\tau)\ .
 \label{mod}
\end{align}
Now the unknowns $\{D_k\}_k$ do not depend on the channel $g$.

\subsubsection{Sphere-torus relation}

By the sphere-torus relation for conformal blocks \eqref{gag}, the modular covariance equations with the spectrum $\mathcal{S}$ \eqref{spec} are equivalent to the crossing symmetry equations for $\left<V_0V_1V_0V_0\right>^\text{sphere}$ with the constraints
\begin{align}
 \mathcal{S}^{(s)} = \mathcal{S}^{(t)}=\mathcal{S}^{(u)}=\mathcal{S}\Big|{}_{\substack{r\in\mathbb{N}\\ rs\in 2\mathbb{Z}}} \quad , \quad D_k^{(s)}=D_k^{(t)}=D_k^{(u)}\ .
 \label{ses}
\end{align}
It may seem that there are many more modular covariance equations, due to the freedom to choose a channel $g\in PSL_2(\mathbb{Z})$. However, it is enough to consider the 3 channels that we called $s,t,u$ \eqref{fstu}, because the corresponding matrices $g=
\left(\begin{smallmatrix} 1 & 0 \\ 0 & 1\end{smallmatrix}\right)$,
$\left(\begin{smallmatrix} 0 & 1 \\ -1 & 0\end{smallmatrix}\right)$, $
\left(\begin{smallmatrix} 1 & -2 \\ 1 & -1\end{smallmatrix}\right)$ generate $PSL_2(\mathbb{Z})$. Modular covariance may also seem stronger because the map $z\mapsto \tau$ \eqref{zqt} is multivalued, so the whole complex $z$-plane corresponds to a small region of the complex $\tau$-plane. However, conformal blocks too are multivalued functions of $z$, and crossing symmetry in the complex $z$-plane implies modular covariance for all $\tau$ by analytic continuation.

In fact, we know the sphere-torus solution not only for the equations, but also for specific solutions, as discussed in Section \ref{sec:scbe}.
In practice, to compute a specific solution, we first have to compute several solutions with larger spectrums  as in Eq. \eqref{sss}. Then the sum of these solutions obeys Eq. \eqref{ses}. In particular, the even spin condition $rs\in 2\mathbb{Z}$ reflects the symmetry under permutations of the 3 identical fields in $\left<V_0V_1V_0V_0\right>^\text{sphere}$.

\subsubsection{Beyond the Moore--Seiberg approach}

Crossing symmetry of sphere 4-point functions, and modular covariance of torus 1-point functions, are only two examples of the conformal bootstrap equations that constrain the spectrum and structure constants. The general equations are obtained by considering an $N$-point function on a Riemann surface of genus $g$: the equations state that all possible decompositions into structure constants and conformal blocks agree.
In the context of rational CFT, Moore and Seiberg have shown that these equations are redundant \cite{ms89b}:
\begin{quote}
 Sphere 4-point functions and torus 1-point functions lead to a complete set of equations.
\end{quote}
And this set is minimal: relaxing modular covariance does give rise to more solutions, for example submodels of minimal models, which exist only on the sphere \cite{bcm24}.

The Moore--Seiberg argument works for a given CFT with a given central charge. However,  the sphere-torus relation changes the central charge, and therefore applies to families of CFTs with different central charges. This is particularly natural when the central charge is a continuous parameter, as in critical loop models and Liouville theory. In the case of Liouville theory, the sphere-torus relation was used for deducing consistency on the torus from consistency on the sphere \cite{hjs10, rs15}.

In an A-series minimal model, crossing symmetry of sphere 4-point functions completely determines 3-point structure constants, but not the spectrum \cite{rib24}. In a 4-point function $\left<V_1V_2V_3V_4\right>^\text{sphere}$, the $s$-channel spectrum is indeed dictated by the fusion rules $V_1\times V_2$ and $V_3\times V_4$, and would therefore be the same in any submodel. On the other hand, in a torus 1-point function $\left<V_1\right>$, the channel spectrum is weakly constrained by fusion rules, leaving it to be determined by modular covariance. Modular covariance gives us access to the full spectrum of the CFT, and eliminates submodels of minimal models. The sphere-torus relation allows us to rephrase the constraints from modular covariance in terms of the 4-point function $\left<V_0V_1V_0V_0\right>^\text{sphere}$.

Consider indeed the torus 1-point function $\left<V^d_{\langle r_1,s_1\rangle}\right>$ in the A-series
minimal model at $\beta^2=\frac{q}{p}$ with $q,r_1,s_1$ odd and $p$ coprime with $q$. By the sphere-torus relation, this is mapped to $\left<V_0V_1V_0V_0\right>^\text{sphere}$ in a CFT with $\beta^2 = \frac{2p}{q}$, with fields of dimensions
\begin{align}
 \Delta_1 = \Delta_{(r_1,\frac{s_1}{2})} = \Delta_{(p-r_1,\frac{q-s_1}{2})}  \quad, \quad \Delta_0 = \Delta_{(0,\frac12)} = \Delta_{(p,\frac{q-1}{2})} \ .
\end{align}
Both dimensions therefore belong to the Kac table. The subtlety is however that both fields cannot be degenerate, otherwise the fusion rules would imply $\left<V_0V_1V_0V_0\right>^\text{sphere}=0$. We assume that $V_1$ is a fully degenerate field of our minimal model, while $V_0$ is not degenerate. Then the channel spectrum of $\left<V_0V_1V_0V_0\right>^\text{sphere}$ is dictated by the fusion product $V_1\times V_0$, while being unconstrained by $V_0\times V_0$. Mapping it back to the torus, this spectrum is
\begin{align}
 \mathcal{S}_{r_1,s_1} = \bigcup_{r=\frac{r_1+1}{2}}^{p-\frac{r_1+1}{2}} \bigcup_{s\overset{2}{=}\frac{s_1+1}{2}}^{q-\frac{s_1+1}{2}} \left\{ V^d_{(r,s)} \right\} \ ,
\end{align}
which coincides with the channel spectrum of $\left<V^d_{\langle r_1,s_1\rangle}\right>$.
The simplest example is the torus partition function $\left<V^d_{\langle 1,1\rangle}\right>$ in the trivial minimal model with $(p, q)=(2, 3)$. This corresponds to fields of dimensions $\Delta_1=\Delta_{(1,1)}$ and $\Delta_0=\Delta_{(2,1)}$ in the Ising minimal model $(p, q)=(4,3)$. Their fusion yields a channel field of dimension $\Delta_{(2,1)}$, which corresponds to a channel field of dimension $\Delta_{(1,1)}$ on the torus. Therefore, the spectrum of the trivial minimal model is encoded in the fusion rules of an extension of the Ising minimal model, which includes the non-degenerate field $V_0$.

This suggests that consistency of CFTs can be reduced to crossing symmetry of sphere 4-point functions, without the need to consider torus 1-point functions. But there are caveats:
\begin{enumerate}
\item The chiral algebra must be the Virasoro algebra: there is no known sphere-torus relation for larger chiral algebras.
 \item It is necessary to consider CFTs at different central charges.
 \item It is necessary to include a field of dimension $\Delta_{(0,\frac12)}$.
 \item In the torus partition function $\left<V^d_{\langle 1,1\rangle}\right>$, multiplicities are integer. This constraint has no counterpart on the sphere.
\end{enumerate}

\section{Results for structure constants}\label{sec:data}

We will now display our results for the polynomials $d_{(r,s)}$ \eqref{eq:15}, for the cases listed in Table \eqref{tab}. Since modular covariance equations are linear, solutions are defined modulo an overall factor, which we fix by a choice of normalization. For maps of the type $(r_1,0,0)$, the spectrum includes diagonal fields, and we set their structure constant to
$d_\text{diag}=1$. For other maps, we fix some other structure constant to $1$.

Since they are unique solutions of the same modular covariance equations, the 1-point functions
$\left<V_{(1,0)}\right>$ and $\left<V_{P_{(1,0)}}\right>$ must coincide. The corresponding polynomials $d_{(r,s)}$ do not quite coincide, because they are normalized differently, and because the reference structure constants $\hat D_{(r,s)}$ and residues $\Theta_1^{r,s}$ differ. This leads to the relation
\begin{align}
 d_{(r,s)}^{\left<V_{P_{(1,0)}}\right>} = wd_{(r,s)}^{\left<V_{(1,0)}\right>} - \delta_{r\in\mathbb{N}^*}\delta_{s\in\mathbb{Z}}\cdot r\Theta^{r,s}_{V_{(1,0)}} \ ,
\end{align}
which is satisfied by our numerical results.

For all 1-point functions with $r_1 \leq 3$, we display all polynomials with $r \leq 3$, except
\begin{itemize}
 \item for $\left\langle V_{(3, \frac23)} \right\rangle$, where we determined only polynomials with $r\leq 2$, as well as $d_{(\frac52,0)}$,
 \item for $\left<V_{(3,0)}\right>$ with the map $(1,1,1)$, where we determined all polynomials with $r\leq 6$ except $d_{(5,\frac15)},d_{(5,\frac25)},d_{(5,\frac35)},d_{(5,\frac45)}$.
\end{itemize}

We use the golden ratio $\varphi = \frac{1+\sqrt{5}}{2}= 2\cos(\frac{\pi}{5})$.

The scripts used to produce the numerical results are available on a public repository \cite{roux26codetorus}. It makes use of the Julia package \texttt{BootstrapVirasoro.jl}, written by one of the authors \cite{roux252}. This package implements the Zamolodchikov recursion to compute Virasoro sphere four-point and torus one-point conformal blocks. The package uses arbitrary-precision arithmetics.
The \TeX\ source of the results is automatically read (parsed) by the Julia code, which checks that the results do satisfy modular covariance.

\subsection{$\left\langle V_{P_1} \right\rangle$ with map $(0, 0, 0)$}
\label{V10-100}

For this 1-point function, $d_{(r, s+1)}(w) = d_{(r, s)}(-w)$ and $d_{(r,s)}=d_{(r,-s)}$.

\begin{subequations}
  \label{eq:1}
  \begin{align}
  \input{results/VP.tex}
  \end{align}
\end{subequations}

\subsection{$\left\langle V_{(1, 0)} \right\rangle$ with map $(1,0,0)$}

For this 1-point function, $d_{(r, s+1)}(w) = -d_{(r, s)}(-w)$ and $d_{(r,s)}=d_{(r,-s)}$.

\begin{subequations}
  \label{eq:2}
  \begin{align}
   \input{results/V10.tex}
  \end{align}
\end{subequations}

\subsection{$\left\langle V_{(2, 0)} \right\rangle$ with maps $(2,0,0)$ and $(1,1,0)$}

\subsubsection{Map $(2, 0, 0)$}

For this 1-point function, $d_{(r, s+1)}(w) = d_{(r, s)}(-w)$ and $d_{(r,s)}=d_{(r,-s)}$.

\begin{subequations}
\label{V20-200}
  \begin{align}
    \input{results/V20_022.tex}
  \end{align}
\end{subequations}

\subsubsection{Map $(1, 1, 0)$}

For this 1-point function, $d_{(r, s+1)} = -d_{(r, s)}$ and $d_{(r,s)}=d_{(r,-s)}$.

\begin{subequations}
  \begin{align}
    \label{eq:3}
    \input{results/V20.tex}
  \end{align}
\end{subequations}

\subsection{$\left\langle V_{(2, 1)} \right\rangle$ with map $(2, 0, 0)$}

For this 1-point function, $d_{(r, s+1)}(w) = d_{(r, s)}(-w)$ and $d_{(r,s)}=d_{(r,-s)}$.

\begin{subequations}
\label{V21-200}
  \begin{align}
   \input{results/V21.tex}
  \end{align}
\end{subequations}

\subsection{$\left\langle V_{(3, 0)} \right\rangle$ with maps $(3,0,0)$ and $(2,1,0)$ and $(1,1,1)$}

\subsubsection{Map $(3, 0, 0)$}

For this 1-point function, $d_{(r, s+1)}(w) = -d_{(r, s)}(-w)$ and $d_{(r,s)}=d_{(r,-s)}$.

\begin{subequations}
  \begin{align}
    \label{eq:4}
\input{results/V30_033.tex}
  \end{align}
\end{subequations}

\subsubsection{Map $(2, 1, 0)$}

For this 1-point function, $d_{(r, s+1)} = -d_{(r, s)}$ and $d_{(r,s)}=d_{(r,-s)}$.

\begin{subequations}
  \label{eq:5}
  \begin{align}
  \input{results/V30_132.tex}
  \end{align}
\end{subequations}

\subsubsection{Map $(1, 1, 1)$}

For this 1-point function, $d_{(r, s+1)} = d_{(r, s)}$ and $d_{(r,s)}=d_{(r,-s)}$. Moreover, we have $r\in\mathbb{N}+\frac12 \implies d_{(r,s)}=0$, allowing us to determine $d_{(r,s)}$ up to $r=6$.

\begin{subequations}
  \label{eq:6}
  \begin{align}
  \input{results/V30.tex}
  \end{align}
\end{subequations}

\subsection{$\big\langle V_{(3, \frac{2}{3})} \big\rangle$ with maps $(3, 0, 0)$ and $(2, 1, 0)$}

\subsubsection{Map $(3, 0, 0)$}

For this 1-point function, $d_{(r, s+1)}(w) = -d_{(r, s)}(-w)$. The parity relation $d_{(r,s)}=d_{(r,-s)}$ does not hold in general: it happens to hold for $(r,s)=(\frac32,\frac23)$, and also for $(r,s)=(2,\frac12)$ as a result of the shift equation, but we numerically find that it fails for $(r,s)=(\frac52,\frac25)$ and $(r,s)=(\frac52,\frac45)$.

\begin{subequations}
  \label{eq:7}
  \begin{align}
  \input{results/V3_23_033.tex}
  \end{align}
\end{subequations}

\subsubsection{Map $(2, 1, 0)$}

For this 1-point function, $d_{(r, s+1)} = -d_{(r, s)}$.

\begin{subequations}
  \label{eq:8}
  \begin{align}
  \input{results/V3_23_132.tex}
  \end{align}
\end{subequations}

\section{Interpretation}

\subsection{Special cases}

Let us focus on 1-point functions of degenerate fields $\left<V^d_{\langle 1,s_1\rangle}\right>$ with $s_1\in\mathbb{N}^*$. The situation depends on the parity of $s_1$:
\begin{itemize}
 \item If $s_1$ is odd, then fusion rules allow the degenerate field $V^d_{\langle r,s\rangle}$ to propagate if $s\geq \frac{s_1+1}{2}$, leading to the vanishing of almost all conformal block residues $R_{r,s}$ \eqref{rrs}:
\begin{align}
 \forall r\in\mathbb{N}^*\ ,\  s\in \frac{s_1+1}{2}+\mathbb{N}\ , \quad R_{r,s} = 0 \ .
 \label{rez}
\end{align}
These vanishings lead to huge simplifications in structure constants and conformal blocks, so that the 1-point function $\left<V^d_{\langle 1,s_1\rangle}\right>$ can be computed exactly, as we will show in the cases $s_1=1,3$.
\item If $s_1$ is even, there are no vanishing residues. Fusion rules allow only finitely many fields to propagate, namely the diagonal fields $V_{(0,s)}$ with $s=(\mathbb{N}+\frac12)\cap [\frac12, \frac{s_1-1}{2}] $. The corresponding 1-point functions involve finitely many conformal blocks, as we will see in the case $s_1=2$.
\end{itemize}

\subsubsection{Partition function of the $O(n)$ model}

We have computed torus 1-point functions of diagonal fields $\left<V_P\right>(w)$, which also depend on the weight $w$ of topologically non-trivial loops. We expect that $Z(w)=\left<V^d_{\langle 1,1\rangle}\right>(w)$ is a $0$-point function, in other words a torus partition function. In fact, the partition function that was originally defined is $Z(n)$, where all loops have the same weight \cite{fsz87}. However, $Z(w)$ should be modular invariant for all $w\in\mathbb{C}$, since the weight $w$ does not depend on which cycle a loop wraps. In CFT language, $Z(w)$ can then be interpreted as a partition function in the presence of a combinatorial defect \cite{rib22}.

It is straightforward to compute $Z(w)$ by the same calculation that was done for $Z(n)$. The result is a combination of the characters
\begin{align}
 \chi_P = \frac{q^{P^2}}{\eta(\tau)}  \ ,
\end{align}
where $\eta(\tau)$ is the Dedekind eta function. The coefficients are written in terms of the Chebyshev polynomials such that $U_d(q+q^{-1})=q^d+q^{-d}$,
\begin{align}
 Z(w) &= \sum_{k\in\mathbb{Z}} \left|\chi_{P_0+k\beta^{-1}}\right|^2 +\sum_{r\in\frac12\mathbb{N}^*}\sum_{s\in\frac{1}{r}\mathbb{Z}}p_{r,s}(w) \chi_{P_{(r,s)}}\overline{\chi}_{P_{(r,-s)}}\ ,
 \label{zw}
 \\ \text{with}\quad
p_{r,s}(w)& = \frac{1}{2r}\sum_{r'=0}^{2r-1} e^{\pi ir's} U_{\text{gcd}(2r,r')}(w)\ ,
\label{prs}
\end{align}
where $w=w(P_0)$ \eqref{wP}.
The polynomials obey $p_{r,-s}=p_{r,s+2}=p_{r,s}$. The first few examples are
\begin{subequations}
\begin{align}
 p_{\frac12, 0}(w) &= w\ ,
 \\
 p_{1,0}(w) &= \tfrac12 (w-1)(w+2)\ ,
 \\
 p_{1,1}(w) &= \tfrac12 (w+1)(w-2)\ ,
 \\
 p_{\frac32, 0}(w) &= \tfrac13 w(w^2-1)\ ,
 \\
 p_{\frac32,\frac23}(w) &= \tfrac13 w(w^2-4)\ ,
 \\
 p_{2, 0}(w) &= \tfrac14 w(w-1)^2(w+2)\ ,
 \\
 p_{2,\frac12}(w) &= \tfrac14 (w^2-1)(w^2-4)\ ,
 \\
 p_{2,1}(w) &= \tfrac14 w(w+1)^2(w-2)\ .
\end{align}
\end{subequations}
We claim that $Z(n)$ agrees with the partition function of the $O(n)$ model \cite{fsz87}. The only subtlety is that the diagonal fields become degenerate, with characters of the type $\chi^d_{\langle r,s\rangle} = \chi_{P_{(r,s)}}-\chi_{P_{(r,-s)}}$. The diagonal term therefore recombines with the non-diagonal with $r=1$ and $s\in 2\mathbb{N}+1$:
\begin{align}
 Z(n) &= \sum_{s\in 2\mathbb{N}+1} \left|\chi^d_{\langle 1,s\rangle}\right|^2 +\sum_{r\in\frac12\mathbb{N}^*}\sum_{s\in\frac{1}{r}\mathbb{Z}}\Big(p_{r,s}(n)+\delta_{r,1}\delta_{s\in2\mathbb{Z}+1}\Big) \chi_{P_{(r,s)}}\overline{\chi}_{P_{(r,-s)}}\ .
\end{align}
Does the partition function $Z(w)$ coincide with our 1-point function $\left<V_{P_{(1,1)}}\right>(w)$? It is not obvious that the field $V_P$ becomes degenerate if $P=P_{(1,1)}$: in Liouville theory, this is a subtle question \cite{rib24}. However, our claim that $\left<V_P\right>(w)$ is the unique solution of modular covariance equations implies
\begin{align}
 \left<V_{P_{(1,1)}}\right>(w)= Z(w)\ .
\end{align}
This amounts to relations between the structure constants and conformal blocks that appear in the decompositions of $\left<V_P\right>(w)$ \eqref{vdg} and $Z(w)$ \eqref{zw}. For conformal blocks, the relation is simply $\mathcal{G}_{(r,s)}= \chi_{P_{(r,s)}}\overline{\chi}_{P_{(r,-s)}}$, as follows from the vanishing of all conformal block residues \eqref{rez}, so that $\mathcal{F}_P = \chi_P$. For the structure constants \eqref{eq:15}, we first notice $\hat{D}_{(r,s)}=1$ by definition, and $\Theta_{1}^{r,s}=0$ because $R_{r,s} = 0$. This leads to the prediction
\begin{align}
 \frac{d_{r,s}(w_1=n,w)}{r\kappa_{r,s}} = p_{r,s}(w)\ .
 \label{w1n}
\end{align}
Our results for $d_{r,s}$ \eqref{eq:1} indeed obey this prediction.

\subsubsection{One-point function of the $O(n)$ model's energy operator $\left<V^d_{\langle 1,3\rangle}\right>$}

The field $V^d_{\left\langle 1, 3 \right\rangle}$ can be interpreted as the energy operator in the $O(n)$
model \cite[(3.51)]{nien84}.
In the case of $\left<V^d_{\langle 1,3\rangle}\right>$, the polynomial factors of the structure constants are the same as for $\left<V^d_{\langle 1,1\rangle}\right>$ \eqref{w1n}, because $w(P_{(1,3)})=w(P_{(1,1)})=n$.
We have checked numerically that this is indeed the case \cite{roux26codetorus}.
Therefore, we know all structure constants analytically. In fact, the reference structure constants \eqref{hdrs} simplify, and may be written in terms of Gamma functions:
\begin{align}
 \hat D_{(r,s)} = \frac{\Gamma(1-\beta^{-2})^2\prod_\pm \Gamma(\beta^{-2}+r\pm \beta^{-2}s)}{\Gamma(\beta^{-2})^2\prod_\pm \Gamma(1-\beta^{-2}+r\pm \beta^{-2}s)} \ .
 \label{hdhd}
\end{align}
Moreover, the residues of conformal blocks obey $R_{r,s}^\text{torus}\neq 0\implies s=1$.
The residues of structure constants all vanish, $\Theta^{r,s}_1=0$. We do not need the non-diagonal sector to cancel poles from the diagonal sector as in Eq. \eqref{rdrd}: the diagonal terms of the 1-point function are already holomorphic in $w$ by themselves, thanks to $\hat D_P$ having a simple zero at $P=P_{(r,1)}$ and a simple pole at $P=P_{(r,-1)}$ for $r\in\mathbb{N}^*$. This leads to the explicit formula
\begin{multline}
 \left<V^d_{\langle 1,3\rangle}\right> = \sum_{P\in P_0+\beta^{-1}\mathbb{Z}} \hat{D}_P \left|\mathcal{F}_P\right|^2
 + \sum_{\substack{r\in\frac12\mathbb{N}^*\ , \  s\in\frac{1}{r}\mathbb{Z}\\ r\notin \mathbb{N}^* \text{ or } |s|\neq 1}}
 \hat{D}_{(r,s)} p_{r,s}(w(P_0)) \mathcal{F}_{P_{(r,s)}}\overline{\mathcal{F}}_{P_{(r,-s)}}
 \\ +\sum_{r=1}^\infty \hat D_{(r,1)} p_{r,1}(w(P_0)) \mathcal{G}_{(r,1)}
 \ ,
\end{multline}
where $p_{r,s}$ is given in Eq. \eqref{prs}, $\hat D_{(r,s)}$ in Eq. \eqref{hdhd}, and $\mathcal{G}_{(r,1)}$ \eqref{ggrs} are the only blocks that are logarithmic.
Similar formulas could be written for $\left<V^d_{\langle 1,s\rangle}\right>$ for any $s\in 2\mathbb{N}+1$, with the same polynomials $p_{r,s}$.

\subsubsection{One-point function of the Potts model's energy operator $\left<V^d_{\langle 1,2\rangle}\right>$}

The degenerate field $V^d_{\left\langle 1, 2 \right\rangle}$ can be interpreted as the energy operator of
the Potts model \cite[above (3.37)]{nien84}. For this field, fusion rules predict that its 1-point function is built from only one conformal block, which must be modular covariant by itself. That block can be computed using the sphere-torus relation, which maps $V^d_{\langle 1,2\rangle}$ to the identity field $V^d_{\langle 1,1\rangle}$. The relevant sphere Virasoro block is simply $\mathcal{F}^{(s),\text{sphere}} = (1-z)^{-\Delta'_{(0,\frac12)}}$, and we deduce the relevant torus Virasoro block $\mathcal{F}_{P_{(0,\frac12)}}$ using Eq. \eqref{faf}. We find
\begin{align}
 \left<V^d_{\langle 1,2\rangle}\right> = \left| \mathcal{F}_{P_{(0,\frac12)}}\right|^2\ , \qquad \text{with} \qquad
 \mathcal{F}_{P_{(0,\frac12)}} = \eta(q)^{2\Delta_{(1,2)}} \ .
 \label{feta}
\end{align}
Modular covariance of our block \eqref{fstu} reduces to the known properties of the eta function \eqref{eta}.
The existence of this one-block solution, and its relevance to the $Q$-state Potts model, were pointed out in \cite{jps19a}: we are now giving a simple formula for the block.

Let us relate this simple 1-point function to our general results. To compute the 1-point function $\left<V_{P_{(1,2)}}\right>$, we set $w_1=-n$ in our general formula \eqref{eq:1}, which does not produce striking simplifications. To compute the 1-point function of the degenerate field $\left<V^d_{\langle 1,2\rangle}\right>$, we must additionally impose the fusion rules for diagonal fields in the spectrum, by setting $w=w(P_{(0,\frac12)})=0$. Then our 1-point function must reduce to the simple solution \eqref{feta} of modular covariance. This implies the vanishings of structure constants
\begin{align}
 \forall s\in\mathbb{N}+\tfrac32 \ , \quad \hat D_{P_{(0,s)}} =0 \qquad , \qquad \forall r\in\tfrac12\mathbb{N}^*\ , s\in \tfrac{1}{r}\mathbb{Z}\ , \quad D_{(r,s)} = 0 \ .
\end{align}
The first vanishing can be straightforwardly checked. The second vanishing is much less trivial, because the reference structure constants $\hat D_{(r,s)}$ are nonzero: it is the other factor of the structure constant \eqref{eq:15} that must vanish, leading to the prediction
\begin{align}
 d_{r,s}\big(w_1=-n,w=0\big) + r\delta_{r\in\mathbb{N}^*}\delta_{s\in\mathbb{Z}} \prod_{j=1-r}^{r-1}2\cos(\pi j \beta^2) = 0 \ .
\end{align}
Our results \eqref{eq:1} satisfy this prediction if $r\in\mathbb{N}+\frac12$, because in this case $d_{(r,s)}(w=0)=0$. For $r\in\mathbb{N}^*$, we have checked that the prediction also holds.

\subsection{Orientable loops in the Potts and $PSU(n)$ models}

In the $Q$-state Potts model and in the $PSU(n)$ loop model \cite{rjrs24}, loops are orientable, and configurations of loops are bicolourable. The corresponding combinatorial maps are also bicolourable. For a 1-point map on the torus $(m_1,m_2,m_3)$, this means $m_1\equiv m_2\equiv m_3\bmod 2$. Among the maps we have considered, $(0, 0, 0), (2, 0, 0)$ and $(1,1,1)$ are bicolourable.

In the critical limit, the Potts and $PSU(n)$ models give rise to CFTs whose non-diagonal fields $V_{(r,s)}$ have $r\in\mathbb{N}^*$: fields with $r\in \mathbb{N}+\frac12$ are absent \cite{rib24}. And indeed, our solution \eqref{eq:6} for the map $(1,1,1)$ obeys $r\in\mathbb{N}+\frac12\implies d_{(r,s)}=0$. For the maps $(0,0,0)$ and $(2,0,0)$, we observe $r\in\mathbb{N}+\frac12\implies d_{(r,s)}(w)=-d_{(r,s)}(-w)$ in our results \eqref{eq:1}, \eqref{V20-200} and \eqref{V21-200}. Bicolourability implies even numbers of topologically nontrivial loops of weight $w$, so the relevant bootstrap solutions are of the type $D_{(r,s)}(w)+D_{(r,s)}(-w)$, which does vanish for $r\in\mathbb{N}+\frac12$.

Therefore, our results are consistent with known properties of orientable loop models, and allow us to compute correlation functions in such models.

\subsubsection{Case of the Potts model}

The spectrum of primary fields of the Potts model is
\begin{align}
  \mathcal{S}^\text{Potts}=
  \left\{V^d_{\langle 1,s\rangle}\right\}_{s\in\mathbb{N}^*} \bigcup \left\{V_{P_{(0,s)}}\right\}_{s\in\mathbb{N}+\frac12}
  \bigcup
 \left\{V_{(r,s)}\right\}_{\substack{r\in \mathbb{N}+2\\ s\in\frac{1}{r}\mathbb{Z}}} \ .
\end{align}
As in the case of the cluster connectivities on the sphere, the corresponding torus 1-point functions can be computed by taking linear combinations of 1-point functions with the spectrum \eqref{spec}, with the 3 values for the channel weight $w\in \{n, -n, 0\}$ that correspond to the degenerate and diagonal fields of the model:
\begin{itemize}
 \item $\left<V^d_{\langle 1,1\rangle}\right>=Z$ is the partition function of the Potts model. It is obtained from the partition function $Z(w)$ \eqref{zw} by
 \begin{align}
  Z= Z(n)+Z(-n)+(n^2-1)Z(0)\ .
  \label{zezzz}
 \end{align}
 The coefficients of $Z(n)$ and $Z(-n)$ must be the same by bicolourability, and the coefficient of $Z(0)$ is determined by the condition that the characters $\chi_{P_{(1,0)}}\overline{\chi}_{P_{(1,0)}}$ do not contribute, by a cancellation between the diagonal contribution of $Z(-n)$ and the non-diagonal contributions of $Z(0),Z(n),Z(-n)$.
 \item $\left<V^d_{\langle 1,s_1\rangle}\right>$ with $s_1\in 2\mathbb{N}+1$ can be calculated just like the partition function, by taking the same linear combination \eqref{zezzz} of solutions of modular covariance, which has a vanishing contribution of $\left|\mathcal{F}_{P_{(1,0)}}\right|^2$. Reference structure constants do not interfere with this combination, because $\hat D_{(1,0)} =\hat D_{P_{(1,0)}}$, as can be checked in the case $s_1=3$ using Eq. \eqref{hdhd}.
 \item $\left<V^d_{\langle 1,2\rangle}\right>$ is the very simple correlation function \eqref{feta}, where fusion rules allow contributions from $w=0$ but not from $w=\pm n$.
 \item $\left<V_{P_{(0,\frac12)}}\right>$ similarly can have contributions from $w=0$ but not from $w=\pm n$. However, $D_{(1,0)}=\frac12 \hat D_{(1,0)} \left[(w-1)(w+2)-\frac{w_1-n}{w+n}\right]$ vanishes for $(w_1,w)=(-n,0)$ but not for $(w_1,w)=(0,0)$. We conclude that the contribution from $w=0$ must vanish, so that $\left<V_{P_{(0,\frac12)}}\right>=0$. And indeed the corresponding operator \cite[(2.12)]{cjv17} in the lattice Potts model with unbroken $S_Q$ symmetry has a vanishing expectation value on any finite graph \cite[Section 3]{cjv17}, and {\em a fortiori} on the torus.
 \item $\left<V_{(2,0)}\right>=\left<V_{(2,1)}\right>=\left<V_{(3,0)}\right>=\left<V_{(3,\frac23)}\right>=0$. For $V_{(3,\frac23)}$, there is no nonzero solution of modular covariance with the map $(1,1,1)$. For $V_{(3,0)}$, the solution with the map $(1,1,1)$ has $D_{(1,0)}\neq 0$. For $V_{(2,0)}$ and $V_{(2,1)}$ with the map $(2,0,0)$, the simple pole of $D_{(1,0)}$ at $w=-n$ and the simple pole of $\hat D_P$ at $P=P_{(1,0)}$ make it impossible to cancel all contributions of $\mathcal{F}_{P_{(1,0)}}$ and $\mathcal{F}'_{P_{(1,0)}}$ by linearly combining the solutions with $w=0,n,-n$. We conjecture that the 1-point functions of all non-diagonal fields vanish. Unlike in the case of $\left<V_{P_{(0,\frac12)}}\right>$, we do not have a lattice interpretation for these vanishings.
\end{itemize}

\subsection{Comparison with lattice models of loops}

The one-point functions corresponding to combinatorial maps can also be considered in lattice models of loops.
Using transfer-matrix techniques we have computed $\langle V_{(1,0)} \rangle$ in the model of \cite{Nienhuis89} defined on $L \times M$ tori,
but other combinatorial maps and related models \cite{nie82,rjrs24} could be treated similarly. The computation is made in a quotient of the periodic dilute Temperley-Lieb algebra on $L$ sites,
in which each non-contractible loop is replaced by the weight $w$, the simultaneous winding of all through-lines is undone \cite{jacobsen15}, and the periodic boundary condition in the $M$-direction
is imposed by a modified Markov trace. In a separate computation, we find the spectrum $\lambda_{(r,s),i}$ of the transfer matrix inside each standard module $(r,s)$ with $2r \le L$ defects and a complex
phase $\mathrm{e}^{i \pi s}$ for each defect that traverses the periodic boundary condition in the $L$-direction. Solving the linear system
\begin{equation}
 \label{examp}
 \langle V_{(1,0)} \rangle_{L,M} = \sum_{(r,s),i} A_{(r,s),i} \left( \lambda_{(r,s),i} \right)^M \,,
\end{equation}
we observe that the ratios between amplitudes for two different values of $w$ are constant on the modules and equal to
\begin{equation}
 \label{ampratio}
 \frac{A_{(r,s),i}(n,w')}{A_{(r,s),i}(n,w)} = \frac{D_{(r,s)}(w')}{\hat{D}_{(r,s)}(w')} \Bigg/ \frac{D_{(r,s)}(w)}{\hat{D}_{(r,s)}(w)} \quad \mbox{independently of } i, \mbox{ for any } r > 0 \,,
\end{equation}
where $D_{(r,s)}$ are as in \eqref{eq:15}.
We further observe that the ratios \eqref{ampratio} are independent of the size $L$, provided that $L \ge 2r$, so that the representation $(r,s)$ exists in finite size.
We have also established that the amplitude ratios are independent of whether $n$ or the local vertex weights are such that the lattice model is critical.
Thus the ratios should be considered algebraic rather than conformal data.

The computation for $L=6$ establishes \eqref{ampratio} for the map $(1,0,0)$, providing an independent check of all the polynomials given in Section~\ref{V10-100}. We hope to justify these observations algebraically and extend them in future work.

\subsection{Fermionic correlation functions}

In the sphere-torus relation \eqref{rel}, the spin of channel fields gets halved when going from the sphere to the torus. We defined critical loop models as containing fields with integer spins, and it may seem that the sphere-torus relation forces us to include fields with half-integer spins on the torus. However, taking the sum \eqref{pzz} of sphere solutions leads to torus solutions with only integer spin fields.

What if we tried to map a single solution to the torus, instead of taking the sum of solutions for different combinatorial maps? We would obtain a fermionic solution of modular covariance, i.e. the channel spectrum would include fields with half-integer spins. That solution would not correspond to a combinatorial map on the torus, and would not be single-valued as a function of the modular parameter $\tau$. Therefore, it would no longer be a purely combinatorial object: to some extent, it would be sensitive to the topology of loops.

This suggests that we could define fermionic loop models, in the same way as there are fermionic minimal models \cite{rw20}. But there is no reason to stop there. Starting from a fermionic solution of crossing symmetry, the sphere-torus relation would produce a solution of modular covariance with spins in $\frac14\mathbb{Z}$. Iterating, we could obtain fields with spins of the type $\frac{1}{2^k}\mathbb{Z}$. But this may be only a hint that arbitrary complex spins are possible, as in spiraling Schramm--Loewner evolutions \cite{hpw25}.

\section*{Acknowledgements}

We are grateful to Max Downing, Rongvoram Nivesvivat and Hubert Saleur for discussions, and for collaboration on related problems. We further thank Max Downing for comments and suggestions on the draft article.
We acknowledge funding from the Agence Nationale de la Recherche through the grant CONFICA (grant No.\ ANR-21-CE40-0003).

\bibliographystyle{../../inputs/morder8}
\bibliography{../../inputs/992}

\end{document}

%% file: results/VP.tex
d_\text{diag} &= 1 \\
d_{(\frac{1}{2}, 0)} &= w \\
d_{(1, 0)} &= (w-1)(w+2) \\
d_{(\frac{3}{2}, 0)} &= w (w_1 - (n+2) w^2+2) \\
d_{(\frac{3}{2}, \frac{2}{3})} &= 2 w (-w_1 + (n-1) w^2-3 n+4) \\
d_{(2, 0)} &= w_1 (-n w^2+2 n+2 w) + (n^2-4) w^4-2 (n^2-6) w^2+2 (n^2-n-4) w-2 n^2 \\
d_{(2, \frac{1}{2})} &= n \Big\{(2-w^2) w_1 + n (w^4-4 w^2+2)\Big\} \\
d_{(\frac{5}{2}, 0)} &= w \Big\{w_1^2 + w_1 \Big[(-n^3-2 n^2+n+2) w^2+2 (n^3+4 n^2-6)\Big] +(n^2+n-2)^2 w^4 \nonumber \\
  & \hspace{-5mm} -2 (2 n^4+4 n^3-7 n^2-9 n+10) w^2+ 7 n^4+10 n^3-28 n^2-24 n+36\Big\} \\
d_{(\frac{5}{2}, \frac{2}{5})} &= 2 w \Big\{\varphi  w_1^2 + w_1 \Big[-w^2 (n^2-1) (n \varphi-\varphi -1) + 2 n^3 \varphi -n^2 (3 \varphi +2) -2 n \varphi \nonumber \\
  & \hspace{-5mm} +n+2 \varphi +3\Big] +w^4 (n^2-n-1) (n^2 \varphi -n-2 \varphi -1) \nonumber \\
  & \hspace{-5mm} +w^2 \Big[-4 n^4 \varphi +4 n^3 (\varphi +1) +14 n^2 \varphi -9 n (\varphi +1)-5 (2 \varphi +1)\Big] \nonumber \\
  & \hspace{-5mm} + 2 n^4 \varphi -n^3 (\varphi +2)-n^2 (11 \varphi +1)+n (6 \varphi +5)+8 \varphi +4\Big\} \\
d_{(\frac{5}{2}, \frac{4}{5})} &= -2 w \Big\{(\varphi -1) w_1^2 +w_1 \Big[-w^2(n^2-1)  (n \varphi -n-\varphi +2) + 2 n^3 (\varphi -1) \nonumber \\
  & \hspace{-5mm} +n^2 (5-3 \varphi ) -2 n \varphi +n+2 \varphi -5\Big] +w^4(n^2-n-1)  (n^2 \varphi -n^2+n-2 \varphi +3) \nonumber \\
  & \hspace{-5mm} +w^2 \Big[-4 n^4 (\varphi -1)+4 n^3 (\varphi -2)+14 n^2 (\varphi -1)-9 n (\varphi -2)-10 \varphi +15\Big] \nonumber \\
  & \hspace{-5mm} + 2 n^4 (\varphi -1)-n^3 (\varphi -3)+n^2 (12-11 \varphi )+n (6 \varphi -11)+8 \varphi -12\Big\} \\
d_{(3, 0)} &= w_1^2 \Big[(n^2-1) w^2-2 n^2+3 w+2\Big] +w_1 \Big[-n (n^4-6 n^2+8) w^4 \nonumber \\
  & \hspace{-5mm} +3 n (n^4-8 n^2+14) w^2-n (n^4-2 n^3-4 n^2+8 n+6) w + 12 n (n^2-3)\Big] \nonumber \\
  & \hspace{-5mm} +(n^2-4)^2 n^2 w^6+(-5 n^6+42 n^4-88 n^2) w^4+(n^2-4)^2 n^2 w^3 \nonumber \\
  & \hspace{-5mm} +(8 n^4-65 n^2+135) n^2 w^2+(-2 n^3+4 n^2+8 n-13) n^2 w -6 n^6+38 n^4-62 n^2 \\
d_{(3, \frac{1}{3})} &= (n^2-1) \Big\{(w^2-2) w_1^2+w_1 \Big[-n (n^2-2) w^4+3 n (n^2-2) w^2+\nonumber \\
  & \hspace{-5mm} (n^3+n^2-3 n-3) w\Big] + 2 n^2+(n^4-4 n^2+3) w^6+(-5 n^4+22 n^2-18) w^4\nonumber \\
  & \hspace{-5mm} +(-n^4+4 n^2-3) w^3 +(5 n^4-27 n^2+24) w^2+ (3 n^4-n^3-13 n^2+3 n+12) w\Big\}

%% file: results/V10.tex
d_\text{diag} &= 1 \\
d_{(\frac{1}{2}, 0)} &= 1 \\
d_{(1, 0)} &= w+1 \\
d_{(\frac{3}{2}, 0)} &= (-n-2) w^2-n+2 \\
d_{(\frac{3}{2}, \frac{2}{3})} &= 2(n-1) w^2-4 (n-2) \\
d_{(2, 0)} &= (n^2-4) w^3+(12-n^2) w+2 (n^2-2 n-4) \\
d_{(2, \frac{1}{2})} &= n^2 w (w^2-3) \\
d_{(\frac{5}{2}, 0)} &= (n^2+n-2)^2 w^4+(-3 n^4-6 n^3+13 n^2+16 n-20) w^2 \nonumber \\
  &\hspace{-5mm} + 5 n^4+2 n^3-27 n^2-12 n+36 \\
d_{(\frac{5}{2}, \frac{2}{5})} &= 2 \Big\{(n^2-n-1)(n^2 \varphi -n-2 \varphi -1) w^4 +\Big[-3 n^4 \varphi +3 n^3 (\varphi +1)+13 n^2 \varphi\nonumber \\
                    &\hspace{-5mm}  -8 n (\varphi +1)-5 (2 \varphi +1)\Big]w^2 + 2 (n-2) \Big[n^2 \varphi -n (2 \varphi +1)-2 \varphi -1\Big] \Big\} \\
d_{(\frac{5}{2}, \frac{4}{5})} &= -2 \Big\{(n^2-n-1)\Big[n^2 (\varphi -1)+n-2 \varphi +3\Big] w^4 \nonumber \\
  &\hspace{-5mm} +\Big[-3 n^4 (\varphi -1)+3 n^3 (\varphi -2) +13 n^2 (\varphi -1)-8 n (\varphi -2)-10 \varphi +15\Big]w^2 \nonumber \\
  &\hspace{-5mm} + 2 (n-2) \Big[n^2 (\varphi -1)+n (3-2 \varphi )-2 \varphi +3\Big]\Big\} \\
d_{(3, 0)} &= n^2 \Big\{(n^2-4)^2 w^5-4 (n^4-9 n^2+20) w^3+(n^2-4)^2 w^2\nonumber \\
  &\hspace{-5mm} +(5 n^4-40 n^2+92) w+n^4-4 n^3+16 n-4\Big\} \\
d_{(3, \frac{1}{3})} &= (n^2-1) \Big\{(n^4-4 n^2+3) w^5+(-4 n^4+20 n^2-18) w^3+(-n^4+4 n^2-3) w^2\nonumber \\
  &\hspace{-5mm} +2 (n^4-10 n^2+12) w+2 (n^4-n^3-5 n^2+3 n+6)\Big\}

%% file: results/V20_022.tex
d_\text{diag} &= 1 \\
d_{(\frac{1}{2}, 0)} &= 0 \\
d_{(1, 0)} &= 1 \\
d_{(\frac{3}{2}, 0)} &= -(n+2) w \\
d_{(\frac{3}{2}, \frac{2}{3})} &= 2 (n-1) w \\
d_{(2, 0)} &= (n-2) \Big\{(n+2) w^2-2)\Big\} \\
d_{(2, \frac{1}{2})} &= n (n w^2-2 n+2) \\
d_{(\frac{5}{2}, 0)} &= (n-1) (n+2) w \Big\{(n^2+n-2) w^2-2 (n^2+n-4)\Big\} \\
 d_{(\frac{5}{2}, \frac{2}{5})} &= 2(2 \varphi +1) w \Big\{- w^2 (n^2-n-1) \Big[n^2 (\varphi -2)+n (2 \varphi -3)+1\Big] \nonumber \\
& \hspace{-5mm} +2 n^4 (\varphi -2)+2 n^3 (\varphi -1)+n^2 (17-9 \varphi )+n (2-3 \varphi )+2 \varphi -5\Big\}\\
 d_{(\frac{5}{2}, \frac{4}{5})} &= 2(2 \varphi -3) w (- w^2 (n^2-n-1) \Big[n^2 (\varphi +1)+2 n \varphi +n-1\Big]\nonumber \\
& \hspace{-5mm} + 2 n^4 (\varphi +1)+2 n^3 \varphi -n^2 (9 \varphi +8)-3 n \varphi +n+2 \varphi +3) \\
d_{(3, 0)} &= (n-2)^2 \Big\{3 n^4+8 n^3+(n+2)^2 n^2 w^4+(n+2)^2 n^2 w-2 n^2\nonumber \\
& \hspace{-5mm} -(3 n^3+12 n^2+10 n-4) n w^2-14 n-1\Big\} \\
 d_{(3, \frac{1}{3})} &= (n-1)(n+1)^2 \Big\{(n^3-n^2-3 n+3) w^4+(-3 n^3+3 n^2+11 n-12) w^2\nonumber \\
& \hspace{-5mm} +(-n^3+n^2+3 n-3) w + 2 (n^2-3 n+2)\Big\}

%% file: results/V20.tex
d_{(\frac{1}{2}, 0)} &= 1 \\
d_{(1, 0)} &= 1 \\
d_{(\frac{3}{2}, 0)} &= -2 (n+1) \\
d_{(\frac{3}{2}, \frac{2}{3})} &= -2 (n-2) \\
d_{(2, 0)} &= 2 (n^2-n-2) \\
d_{(2, \frac{1}{2})} &= 0 \\
d_{(\frac{5}{2}, 0)} &= 2 (2 n^4+2 n^3-7 n^2-4 n+8) \\
d_{(\frac{5}{2}, \frac{2}{5})} &= -2 (n^2-n-2) \Big\{n^2 \varphi -n (\varphi +1)-1\Big\} \\
d_{(\frac{5}{2}, \frac{4}{5})} &= 2 (n^2-n-2) \Big\{n^2 (\varphi -1)-n (\varphi -2)+1\Big\} \\
d_{(3, 0)} &= (n-2)^2 (2 n^4+6 n^3+7 n^2+6 n+3) \\
d_{(3, \frac{1}{3})} &= (n+1)^2 (n^4-3 n^3-n^2+9 n-6)

%% file: results/V21.tex
d_\text{diag} &= 1 \\
d_{(\frac12, 0)} &= 0 \\
d_{(1, 0)} &= -1 \\
d_{(\frac{3}{2}, 0)} &= (n+2) w \\
d_{(\frac{3}{2}, \frac{2}{3})} &= -2 (n-1) w \\
d_{(2, 0)} &= -(n+2) \Big\{(n-2) w^2+2\Big\} \\
d_{(2, \frac{1}{2})} &= -n \Big\{n w^2-2 (n+1)\Big\} \\
d_{(\frac{5}{2}, 0)} &= -(n^2+n-2)^2 w (w^2-2) \\
d_{(\frac{5}{2}, \frac{2}{5})} &= -2 w \Big\{(n^2-n-1)  (n^2 \varphi -n-2 \varphi -1)w^2-2 n^4 \varphi +2 n^3 (\varphi +1)\nonumber \\
& \hspace{-5mm} +n^2 (9 \varphi +1)-n (5 \varphi +6)-8 \varphi -5\Big\} \\
d_{(\frac{5}{2}, \frac{4}{5})} &= 2 w \Big\{(n^2-n-1)  \Big[n^2 (\varphi -1)+n-2 \varphi +3\Big]w^2-2 n^4 (\varphi -1)\nonumber \\
& \hspace{-5mm} +2 n^3 (\varphi -2)+n^2 (9 \varphi -10) +n (11-5 \varphi )-8 \varphi +13\Big\} \\
d_{(3, 0)} &= -(n+2)^2 \Big\{(n-2)^2 n^2 w^4-2 n^2-(3 n^3-12 n^2+10 n+4) n w^2\nonumber \\
& \hspace{-5mm} +(n-2)^2 n^2 w+3 n^4-8 n^3+14 n-1\Big\} \\
d_{(3, \frac{1}{3})} &= -(n-1)^2 (n+1) \Big\{(n^3+n^2-3 n-3) w^4+(-3 n^3-3 n^2+11 n+12) w^2\nonumber \\
& \hspace{-5mm} +(-n^3-n^2+3 n+3) w-2 (n^2+3 n+2)\Big\}

%% file: results/V30_033.tex
d_\text{diag} &= 1 \\
d_{(\frac{1}{2}, 0)} &= 0 \\
d_{(1, 0)} &= 0 \\
d_{(\frac{3}{2}, 0)} &= -n-2 \\
d_{(\frac{3}{2}, \frac{2}{3})} &= 2 (n-1) \\
d_{(2, 0)} &= (n^2-4) w \\
d_{(2, \frac{1}{2})} &= n^2 w \\
d_{(\frac{5}{2}, 0)} &= (n-1) (n+2) \Big\{(n^2+n-2) w^2-n^2-3 n+2\Big\} \\
d_{(\frac{5}{2}, \frac{2}{5})} &= 2(2 \varphi +1) \Big\{- (n^2-n-1)  \Big[n^2 (\varphi -2)+n (2 \varphi -3)+1\Big]w^2\nonumber \\
   & \hspace{-5mm} +n^4 (\varphi -2)+2 n^3 (\varphi -1)+n^2 (9-5 \varphi )-4 n (\varphi -1)-1\Big\} \\
d_{(\frac{5}{2}, \frac{4}{5})} &= 2(2 \varphi -3) \Big\{- (n^2-n-1)  \Big[n^2 (\varphi +1)+2 n \varphi +n-1\Big]w^2\nonumber \\
   & \hspace{-5mm} + n^4 (\varphi +1)+2 n^3 \varphi - n^2 (5 \varphi +4)-4 n \varphi +1\Big\} \\
d_{(3, 0)} &= n^2 (n^2-4) \Big\{(n^2-4) w^3-2 (n^2-6) w+n^2-4\Big\} \\
d_{(3, \frac{1}{3})} &= (n^2-1)\Big\{(n^4-4 n^2+3) w^3+(-2 n^4+9 n^2-6) w-n^4+4 n^2-3\Big\}

%% file: results/V30_132.tex
d_{(\frac{1}{2}, 0)} &= 1 \\
d_{(1, 0)} &= 1 \\
d_{(\frac{3}{2}, 0)} &= -2n-3 \\
d_{(\frac{3}{2}, \frac{2}{3})} &= -2n \\
d_{(2, 0)} &= 2(n^2-n-4) \\
d_{(2, \frac{1}{2})} &= 0 \\
d_{(\frac{5}{2}, 0)} &= 4 n^4+6 n^3-10 n^2-9 n+10 \\
d_{(3, 0)} &= 2n^2 (n^4-2 n^3-7 n^2+8 n+15) \\
d_{(3, \frac{1}{3})} &= n(n-1)(n^2-3)(n+1)^2

%% file: results/V30.tex
d_{(1, 0)} &= 1\\
d_{(2, 0)} &= n^2 \\
d_{(2, \frac12)} &= -n^2 \\
d_{(3, 0)} &= 2 n^2 (11-6 n^2+n^4) \\
d_{(3, \frac13)} &= -n^2 (2-3 n^2+n^4) \\
d_{(4, 0)} &= n^4 (-2+n^2) (-36+46 n^2-17 n^4+2 n^6)\\
d_{(4, \frac14)} &= -n^2 (n^2-2)^2 (2-10 n^2+3 n^4)\\
d_{(4, \frac12)} &= -n^4 (n^2-2)^2 (16-11 n^2+2 n^4)\\
d_{(5, 0)} &= 2 n^4 (-2+n^2)^2 (2 n^{12} - 32 n^{10} + 205 n^8 - 662n^6 + 1110n^4 -892n^2 + 277) \\
d_{(6, 0)} &= n^6 (n^2-2)^2 (2 n^{20}-46 n^{18}+459 n^{16}-2608 n^{14}+9326 n^{12}-21946 n^{10}\nonumber\\
& \hspace{-5mm} +34573 n^8-36228 n^6+24096 n^4-8920 n^2+1388)\\
d_{(6, \frac16)} &= n^2 (n^2-3) (n^2-2) (n^{24}-27 n^{22}+313 n^{20}-2054 n^{18}+8445 n^{16}\nonumber\\
& \hspace{-5mm} -22694 n^{14}+40294 n^{12}-46678 n^{10}+34053 n^8-14731 n^6\nonumber\\
& \hspace{-5mm} +3471 n^4-384 n^2+15)\\
d_{(6, \frac13)} &= -(n-1) n^2 (n+1) (n^2-3)^2 (n^2-2) (n^{20}-15 n^{18}+88 n^{16}-241 n^{14} \nonumber\\
& \hspace{-5mm} +229 n^{12}+333 n^{10}-1065 n^8+1025 n^6-394 n^4+42 n^2-1)\\
d_{(6, \frac12)} &= -n^6 (n^2-2)^2 (n^2-3)^2 (2 n^{16}-38 n^{14}+301 n^{12}-1280 n^{10}+3149 n^8\nonumber\\
& \hspace{-5mm} -4524 n^6+3668 n^4-1528 n^2+256)

%% file: results/V3_23_033.tex
d_\text{diag} &= 1 \\
d_{(\frac{1}{2}, 0)} &= 0 \\
d_{(1, 0)} &= 0 \\
d_{(\frac{3}{2}, 0)} &= n+2 \\
d_{(\frac{3}{2}, \frac{2}{3})} &= 2 (1-n) \\
d_{(\frac{3}{2}, -\frac{2}{3})} &= 2 (1-n) \label{eq:drms}\\
d_{(2, 0)} &= -(n^2-4) w \\
d_{(2, \frac{1}{2})} &= -n^2 w \\
d_{(\frac{5}{2}, 0)} &= (1-n) (n+2) \Big\{(n^2+n-2) w^2-n^2+5\Big\}

%% file: results/V3_23_132.tex
d_{(\frac{1}{2}, 0)} &= 1 \\
d_{(1, 0)} &= -1 \\
d_{(\frac{3}{2}, 0)} &= n+3 \\
d_{(\frac{3}{2}, \frac{2}{3})} &= -2n \\
d_{(\frac{3}{2}, -\frac{2}{3})} &= 2(2n-3) \\
d_{(2, 0)} &= -n^2+n+4 \\
d_{(2, \frac{1}{2})} &= -\sqrt{3} (n-1) n \\
d_{(\frac{5}{2}, 0)} &= -2 n^4-3 n^3+8 n^2+6 n-11